\documentclass[superscriptaddress,aps,pra,twocolumn,showpacs,nofootinbib,longbibliography]{revtex4-1}
\usepackage{amsmath}
\usepackage{latexsym}
\usepackage{amssymb}
\usepackage[colorlinks=true,citecolor=blue,urlcolor=blue]{hyperref}
\usepackage{color}
\usepackage{graphics,epstopdf}
\usepackage{soul}
\usepackage[demo]{graphicx}
\usepackage{subcaption}
\usepackage{lipsum}

\begin{document}

\title{Preservation of quantum non-bilocal correlations in noisy entanglement-swapping experiments
using weak measurements}

\author{Shashank Gupta}
\email{shashankg687@bose.res.in}
\affiliation{S. N. Bose National Centre for Basic Sciences, Block JD, Sector III, Salt Lake, Kolkata 700 098, India}

\author{Shounak Datta}
\email{shounak.datta@bose.res.in}
\affiliation{S. N. Bose National Centre for Basic Sciences, Block JD, Sector III, Salt Lake, Kolkata 700 098, India}

\author{A. S. Majumdar}
\email{archan@bose.res.in}
\affiliation{S. N. Bose National Centre for Basic Sciences, Block JD, Sector III, Salt Lake, Kolkata 700 098, India}


\date{\today}

\begin{abstract}
A tripartite quantum network is said to be bilocal if two independent sources produce a pair of bipartite entangled states. Quantum non-bilocal correlation emerges when the central party which possesses two particles from two different sources performs Bell-state measurement on them and  nonlocality is generated between the other two uncorrelated systems in this entanglement-swapping protocol. The interaction of such systems with the environment  reduces quantum non-bilocal correlations. Here we show that the diminishing effect modelled by the amplitude damping channel  can be slowed down by employing the technique of weak measurements and reversals. It is demonstrated that
for a large range of parameters the quantum non-bilocal correlations are preserved against decoherence
by taking into account the average success rate of the post-selection governing weak measurements.
\end{abstract}


\maketitle

\section{Introduction} \label{1}

It is well known since the formulation of the Einstein-Podolsky-Rosen(EPR) paradox\cite{EPR} that classical intuition encounters difficulties in explaining experimental facts in the quantum domain. Correlations in measurement outcomes of two or more distant parties refute the existence of local realist theories leading to nonlocality\cite{Bell} as a feature of quantum theory. Nonlocality is used as resource in quantum information processing in many examples, such as quantum key distribution (QKD)\cite{QKD1,QKD2}, random number generation\cite{RNG}, device-independent way of witnessing entanglement\cite{EW}, etc. Bell's inequality\cite{Bell} provides an upper bound to correlations in a local realist model where the local hidden variable $\lambda$ corresponds to a specific source defining the state of the system.

\begin{figure}[!ht]
\resizebox{5.5cm}{5.5cm}{\includegraphics{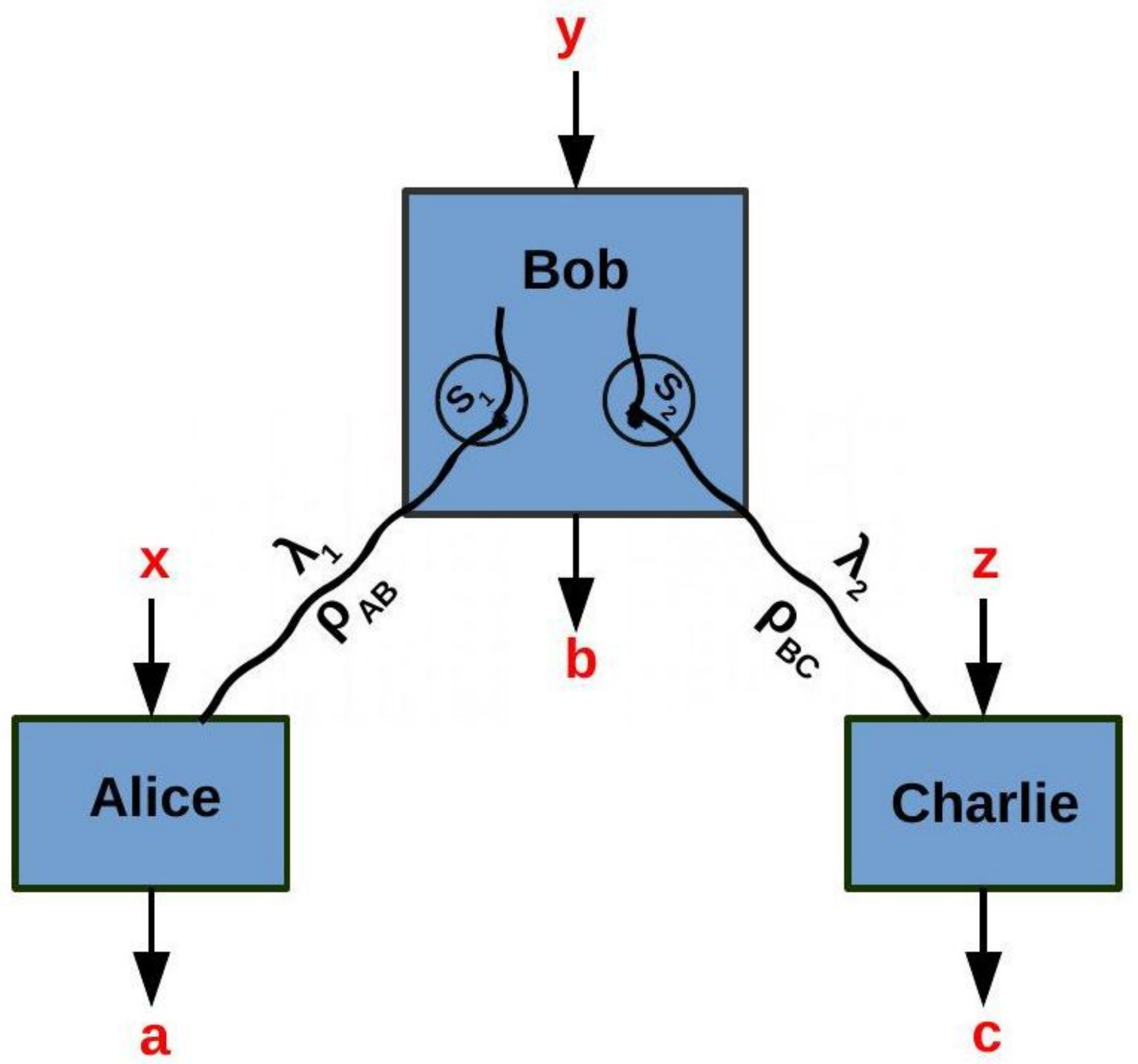}}
\caption{\footnotesize Schematic diagram of the bilocal scenario}
\label{model}
\end{figure}

The Bell scenario can be recast in tripartite network where two entangled pairs coming from two different sources are distributed in such a way, that the central party occupies one particle from each of the two sources. Next, a joint measurement, {\it viz.} Bell-state measurement (BSM) by the central party is used to generate entanglement between two initially uncorrelated distant parties. This is a very familiar scheme of entanglement swapping\cite{swap} generally used in quantum teleportation protocols. Construction of a model through the assumption of independent sources $S_1$ and $S_2$ producing two bipartite entangled states, $\rho_{AB}$ and $\rho_{BC}$ governed by two independent set of hidden variables $\lambda_1$ and $\lambda_2$ constitute the framework of bilocality\cite{cyril,cyril1} (Fig.\ref{model}). Violation of a bilocality inequality based on bilocal realist assumptions suffices to show quantumness through generation of nonlocality between two uncorrelated systems ($A$ and $C$). 

Quantum non-bilocal correlations are those which originate from the outcomes of measurements performed by three parties where the central party performs BSM and other two parties do dichotomic projective qubit measurements randomly as per local quantum random number generators(QRNG)\cite{RNG1}. Instead of choosing a single tripartite state, here one assumes that $\rho_{AB}$ is shared between Alice and Bob whereas $\rho_{BC}$ is shared between Bob and Charlie. Bilocal states form a strict subset of local states\cite{cyril1} and hence leave a possibility for a set of states which are non-bilocal but local. In the context of experiments, bilocality by means of independent sources demands less complexity and reduces the blending of noise with the intermediate group of particles in quantum repeater networks\cite{rep}. Such studies have been recently extended to $n$-locality networks\cite{star1,star2} and also experimentally verified in \cite{expt1,expt2}.

Practical applications of quantum correlations  face hindrance due to interaction with the ubiquitous environment. The effect of environment generally decreases correlations\cite{nielsen}, apart from exceptions in certain special contexts\cite{except1,except2,except3}. It is thus imperative to protect such correlations against decoherence. A promising avenue in this direction is provided by weak measurements\cite{vaidman} which are based on weak coupling between the system and the measurement device, and have interesting applications in different arenas such as demonstration of spin Hall effect\cite{hall} and wave-particle duality\cite{cavity}. Weak measurements have been employed to protect quantumness of bipartite systems used in several applications such as generation of entanglement\cite{weak1}, enhancement of teleportation fidelity\cite{weak2}, and protection of the secret key rate in one-sided device independent QKD\cite{weak3}. The technique of  weak measurement and its reversal has been used to preserve correlations in quantum states transmitting through amplitude damping channels (ADC).

In the present work, our motivation is to explore whether non-bilocal quantum correlations generated in entanglement swapping experiments can be protected against amplitude damping using weak measurements. Here we consider the following set-up. Two independent sources at Bob's lab produce a pair of maximally entangled pure states. From each pair, Bob sends one particle to Alice and Charlie respectively, through an environment modelled by ADC. We first show that quantum non-bilocal correlations are non-increasing when one or both of the end particles are affected by environment for different kinds of joint state measurements. We next apply the technique of weak measurement  and its reversal and show that the magnitude of violation of bilocality inequalities under ADC  can be enhanced. We finally consider the average effect of success and failure of post-selection ingrained in weak measurements on bilocal correlations and demonstrate that the average correlations using the weak measurement technique is improved over a range of parameters.

The paper is organised in the following way. In Sec. \ref{2}, we recapitulate the characterization of bilocal correlations. In Sec. \ref{3}, we discuss briefly the technique of weak and reverse weak measurement in the presence of ADC. In Sec. \ref{4}, we evaluate the effect of decoherence on bilocal correlations under three set-ups of BSMs. In Sec. \ref{5}, we demonstrate the preservation of non-bilocal correlations in the framework of the weak measurement technique. Subsequently, in Sec. \ref{6}, we compute the average effect of weak measurement on the non-bilocal correlation dissipated under the action of ADC. Finally, we conclude this paper with a summary in Sec. \ref{7}.

\section{Overview of bilocal correlations} \label{2}

Let us consider that three parties, Alice, Bob and Charlie perform three black box measurements $x,y,z$, respectively.  Bell nonlocality is manifested if the joint probability, $p(a,b,c|x,y,z)$ of having outputs $a,b,c$ corresponding to inputs $x,y,z$ respectively, cannot be decomposed as
\begin{equation}
p(a,b,c|x,y,z) = \sum_{\lambda} p(\lambda) ~ p(a|x,\lambda) ~ p(b|y,\lambda) ~ p(c|z,\lambda)
\label{bell}
\end{equation}
where the local hidden variable $\lambda$ is distributed with probability $p(\lambda)$ satisfying $\sum_{\lambda} p(\lambda)=1$. $p(a|x,\lambda)$ is the conditional probability of getting outcome $a$ for input $x$ depending upon the hidden variable $\lambda$ at Alice's side. Similarly, the conditional probabilities $p(b|y,\lambda)$ and $p(c|z,\lambda)$ correspond to Bob's output $b$ from input $y$ and Charlie's output $c$ from input $z$ respectively, contingent upon the hidden variable $\lambda$.

If instead of choosing a single source for distributing $\lambda$, it is assumed that two sources $S_1$ and $S_2$ generate two sets of distributions of hidden variables, say, $\lambda_1$ and $\lambda_2$, the local realist description (see Fig.\ref{model}) can also be implemented by the following decomposition:
\begin{align}
p(a,b,c|x,y,z) = \sum_{\lambda_1,\lambda_2} & p(\lambda_1,\lambda_2) ~ p(a|x,\lambda_1) \nonumber\\
& \times p(b|y,\lambda_1,\lambda_2) ~ p(c|z,\lambda_2)
\label{bilocal}
\end{align}
where, $\lambda_1=\lambda_2=\lambda$ reproduces Eq.(\ref{bell}).

Eq.(\ref{bilocal}) together with the assumption of factorizability in the distribution of $\lambda_1$ and $\lambda_2$, i.e., $p(\lambda_1,\lambda_2)=p(\lambda_1) p(\lambda_2)$ constitute the assumption of bilocality\cite{cyril1}. The factorizability assumption reflects the independence of the sources $S_1$ and $S_2$ from each other. Therefore, the distributions of $\lambda_1$ and $\lambda_2$ separately satisfy the normalizability conditions: $\sum_{\lambda_1} p(\lambda_1)=\sum_{\lambda_2} p(\lambda_2)=1$. The above bilocal description enables one to construct bilocality inequalities. It follows that all bilocal states must be local whereas the counter-statement is not always true. Further, bilocal states are not convex though all local deterministic boxes form convex hull of
the bilocal set \cite{cyril1}.

From alternative formulations of the bilocality assumption which are useful for practical purposes, non-linear bilocal inequalities can be constructed \cite{cyril1}, the violation of which imply the presence of non-bilocality that can be detected in experiments. Depending on different kinds of Bell-state measurement performed by Bob, such inequalities can be classified into the following types. Here
 we consider no  imperfection in the detectors used by Alice, Bob and Charlie.

\subsection{Scenario of binary inputs and outputs}

Consider the scenario where Alice, Bob and Charlie have binary inputs $x,y,z\in\lbrace0,1\rbrace$ corresponding to measurements $A_x,B_y,C_z$ respectively, and binary outputs $a,b,c\in\lbrace0,1\rbrace$ for each input. A tripartite correlation in terms of the joint probability distribution denoted by $P^{22}$ can be written as,
\begin{align*}
\langle A_x B_y C_z \rangle_{P^{22}} = \sum_{a,b,c} (-1)^{a+b+c} P^{22}(a,b,c|x,y,z)
\end{align*}
Now, taking suitable linear combinations one may define,
\begin{align*}
I^{22} = \frac{1}{4} \sum_{x,z=0,1} \langle A_x B_0 C_z \rangle_{P^{22}}
\end{align*}
and
\begin{align*}
J^{22} = \frac{1}{4} \sum_{x,z=0,1} (-1)^{x+z} \langle A_x B_1 C_z \rangle_{P^{22}}
\end{align*}
The sufficient condition for the correlation $P^{22}$ to be bilocal is given by~\cite{cyril1},
\begin{equation}
B^{22} = \sqrt{|I^{22}|}+\sqrt{|J^{22}|} \leq 1
\label{2-2}
\end{equation}
Here, $A_x$ and $C_z$ are dichotomic qubit measurements, whereas $B_y$ is a complete BSM performed by Bob when he wants to distinguish either $\lbrace|\phi^{\pm}\rangle\rbrace$ vs $\lbrace|\psi^{\pm}\rangle\rbrace$ Bell states, or $\lbrace|\phi^{+}\rangle,|\psi^{+}\rangle\rbrace$ vs $\lbrace|\phi^{-}\rangle,|\psi^{-}\rangle\rbrace$ Bell states where $|\phi^{\pm}\rangle = \frac{1}{\sqrt{2}} (|00\rangle \pm |11\rangle)$ and $|\psi^{\pm}\rangle = \frac{1}{\sqrt{2}} (|01\rangle \pm |10\rangle)$ with $\lbrace |0\rangle,|1\rangle \rbrace$ as eigenstates of $\sigma_z$. Thus, $B_0 = |\phi^{+}\rangle\langle\phi^{+}| + |\phi^{-}\rangle\langle\phi^{-}| - |\psi^{+}\rangle\langle\psi^{+}| - |\psi^{-}\rangle\langle\psi^{-}|$ reduces to $\sigma_z \otimes \sigma_z$ and $B_1 = |\phi^{+}\rangle\langle\phi^{+}| - |\phi^{-}\rangle\langle\phi^{-}| + |\psi^{+}\rangle\langle\psi^{+}| - |\psi^{-}\rangle\langle\psi^{-}|$ reduces to $\sigma_x \otimes \sigma_x$.

As the bilocal assumptions remain equivalent under a larger class of separable measurements\cite{andreoli}, hence we can choose $A_x\equiv \vec{a}_x.\vec{\sigma}$, $B_y\equiv \vec{b}_y^A.\vec{\sigma} \otimes \vec{b}_y^C.\vec{\sigma}$ and $C_z\equiv \vec{c}_z.\vec{\sigma}$ where $\vec{\sigma}=(\sigma_x,\sigma_y,\sigma_z)$ and $x,y,z\in\lbrace0,1\rbrace$. For a general quantum state $\rho_{AB} \otimes \rho_{BC}$ with local projective measurements, the bilocality parameter $B^{22}$ turns out to be
\begin{align}
B^{22} = & \frac{1}{2} \Big(\sqrt{|(\vec{a}_0 + \vec{a}_1).T_{\rho_{AB}} \vec{b}_0^A| ~ |\vec{b}_0^C. T_{\rho_{BC}} (\vec{c}_0 + \vec{c}_1)|} \nonumber\\
& + \sqrt{|(\vec{a}_0 - \vec{a}_1).T_{\rho_{AB}} \vec{b}_1^A| ~ |\vec{b}_1^C. T_{\rho_{BC}} (\vec{c}_0 - \vec{c}_1)|}\Big)
\label{andr}
\end{align}
where, $T_{\rho_{AB(BC)}}$ is a $3\times 3$ correlation matrix formed with the elements given by, $T_{\rho_{AB(BC)}}^{ij} = \operatorname{Tr} [\rho_{AB(BC)} \sigma_i \otimes \sigma_j] ~ \forall ~ i,j\in\lbrace x,y,z \rbrace$. Maximizing the bilocality parameter w.r.t. all measurement settings, we have
\begin{equation}
B_{max}^{22} = \sqrt{\sqrt{t_1^A t_1^C} + \sqrt{t_2^A t_2^C}}
\label{andrmax}
\end{equation}
where, $t_1^A$ and $t_2^A$ ($t_1^C$ and $t_2^C$) are the two greater (positive) eigenvalues of the matrix $T_{\rho_{AB}}^t T_{\rho_{AB}}$ ($T_{\rho_{BC}}^t T_{\rho_{BC}}$) with $t_1^A \geq t_2^A$ ($t_1^C \geq t_2^C$). $T_{\rho_{AB(BC)}}^t$ are the transposition matrices of $T_{\rho_{AB(BC)}}$.

\subsection{Scenario of one input and four outputs for Bob}

Alice and Charlie have binary inputs and binary outputs similar to the previous case.  Bob performs a complete BSM with four distinguishable outputs $b=b_0 b_1=\lbrace 00,01,10,11 \rbrace$.  Corresponding to the joint probability distribution $P^{14}$, Alice, Bob and Charlie can share measurement correlations in the following way:
\begin{align*}
\langle A_x B_y C_z \rangle_{P^{14}} = \sum_{a,b,c} (-1)^{a+b_y+c} P^{14}(a,b_0 b_1,c|x,z)
\end{align*}
Using linear combination of the correlators, let us denote
\begin{align*}
I^{14} = \frac{1}{4} \sum_{x,z=0,1} \langle A_x B_0 C_z \rangle_{P^{14}}
\end{align*}
and
\begin{align*}
J^{14} = \frac{1}{4} \sum_{x,z=0,1} (-1)^{x+z} \langle A_x B_1 C_z \rangle_{P^{14}}
\end{align*}
$P^{14}$ is bilocal if the following necessary condition holds\cite{cyril1}
\begin{equation}
B^{14} = \sqrt{|I^{14}|}+\sqrt{|J^{14}|} \leq 1
\label{1-4}
\end{equation}

\subsection{Scenario of one input and three outputs for Bob}

Comparing with the previous scenario, Bob can distinguish among three outputs here in the context of complete BSM, i.e. $b=b_0 b_1=00$ or $01$ or $\lbrace 10$ or $11 \rbrace$. Bob is unable to distinguish between the outputs $10$ and $11$. Then, corresponding to the joint probability distribution $P^{13}$, tripartite correlators can be defined as,
\begin{align*}
\langle A_x B_0 C_z \rangle_{P^{13}} = & \sum_{a,c} (-1)^{a+c} [P^{13}(a,00,c|x,z) \\
& +P^{13}(a,01,c|x,z) \\
& -P^{13}(a,\lbrace 10 ~ or ~ 11 \rbrace,c|x,z)]
\end{align*}
and
\begin{align*}
\langle A_x B_1 C_z \rangle_{P^{13}} = & \sum_{a,c} (-1)^{a+c} [P^{13}(a,00,c|x,z)\\
& -P^{13}(a,01,c|x,z)]
\end{align*}
Linear combination can be represented as,
\begin{align*}
I^{13} = \frac{1}{4} \sum_{x,z=0,1} \langle A_x B_0 C_z \rangle_{P^{13}}
\end{align*}
and
\begin{align*}
J^{13} = \frac{1}{4} \sum_{x,z=0,1} (-1)^{x+z} \langle A_x B_1 C_z \rangle_{P^{13},b_0=0}
\end{align*}
Bilocality of correlation $P^{13}$ can be necessarily revealed if\cite{cyril1}
\begin{equation}
B^{13} = \sqrt{|I^{13}|}+\sqrt{|J^{13}|} \leq 1
\label{1-3}
\end{equation}
The  distinction between the violation of bilocality inequality and Bell-CHSH inequality in QM has been analysed in details in Refs.\cite{gisin,andreoli}.

\section{Technique of weak measurement in the presence of decoherence} \label{3}

The amplitude damping channel(ADC)\cite{nielsen} models 
a system dissipating energy to the environment by spontaneously emitting photons. It is a completely positive trace preserving(CPTP) map which acts on a two-level system, say $\varrho$, in the following way:
\begin{equation}
\varrho \rightarrow \varrho' = \varepsilon(\varrho) = \sum_{i=0}^1 K_i \varrho K_i^{\dagger}
\end{equation}
where, Kraus operators represented by
\begin{equation}
K_0 = \begin{pmatrix}
1 & 0\\
0 & \sqrt{1-p}\\
\end{pmatrix}; ~~~ K_1 = \begin{pmatrix}
0 & \sqrt{p}\\
0 & 0\\
\end{pmatrix}
\label{damp}
\end{equation}
satisfy completeness relation $\sum_{i=0}^1 K_i^{\dagger} K_i = \openone_2$. The decoherence parameter $p$ ($0\leq p \leq 1$) relates to the probability of photon emission by the system. Here, the Kraus operators $K_i$ ($i\in\lbrace0,1\rbrace$) are expressed in the eigenbasis of $\sigma_z$. In the  course of our calculation, we consider the decoherence strength to be same for Alice and Charlie, for the sake of simplicity.

Weak measurement signifies weak interaction between the system and the detector\cite{weak1}. If the detector clicks,  any residual entanglement remaining in the system with another system is completely destroyed. Otherwise, the coupling will persist. Let us consider that the measurement device detects the system with probability $w$, when the system is in the state $|1\rangle$. Then, the evolution of the system is governed by the following Kraus operator in the eigenbasis of $\sigma_z$:
\begin{equation}
W_0 = \begin{pmatrix}
0 & 0\\
0 & \sqrt{w}\\
\end{pmatrix}
\end{equation}
When the system is not detected by the device, the Kraus operator representing the evolution of the system can be written as
\begin{equation}
W_1 = \begin{pmatrix}
1 & 0\\
0 & \sqrt{1-w}\\
\end{pmatrix}
\label{weak}
\end{equation}
where $w$ ($0\leq w \leq 1$) is regarded as the strength of weak measurement.
It can be easily checked that the matrix $W_0$ has no inverse, and thus the operation of detection is an irreversible process. On the other hand, $W_1$ is reversible and enables restoring the system to its initial state. In the following analysis, we will associate a probability   with  successful and unsuccessful detection, and consider further operations on the  the system when it is not detected under weak measurement.

Weak measurement is performed  prior to the decoherence effect.  Then, subsequent to the action of the ADC, reverse weak measurement is performed. This constitutes the technique of weak measurement\cite{wktch1,wktch2,wktch3,weak1,wktch4,wktch5,wktch6}. Reverse weak measurement corresponding to the scenario where the system is not detected (i.e. represented by $W_1$), has the following Kraus operator in $\sigma_z$-eigenbasis:
\begin{equation}
R_1 = \begin{pmatrix}
\sqrt{1-r} & 0\\
0 & 1\\
\end{pmatrix}
\label{reverse}
\end{equation}
where, $r$ ($0\leq r \leq 1$) is the strength of reverse weak measurement which can be optimized
depending on the requirement of the situation.

\section{Bilocal correlations in the presence of decoherence} \label{4}

We are interested in a bilocal scenario where Bob holds two independent sources which individually produce two bipartite entangled states associated with the hidden variables $\lambda_1$ and $\lambda_2$, respectively.  Bob keeps one particle from each pair in his possession and sends the other particle from each pair to Alice and Charlie, respectively, who are located far apart from Bob and also from each other. We now  consider two cases:
\begin{enumerate}
\item \label{c1} Alice's particle is transmitted through ADC and Charlie's particle through an ideal noiseless channel;
\item \label{c2} Both Alice's and Charlie's particles are transmitted through unconnected ADCs, but with equal strength.
\end{enumerate}
After setting-up the above two situations, Alice, Bob and Charlie perform entanglement-swapping experiments to come up with quantum non-bilocal correlations among them.

In this section, we discuss the two cases-\ref{c1} and -\ref{c2} separately. To evaluate the effect of decoherence on quantum non-bilocal correlations, we start with identical maximally entangled bipartite states shared by both the Alice-Bob pair and the Bob-Charlie pair. Two classes of such Bell states are given by,
\begin{eqnarray}
|\psi^{\pm}\rangle_{AB(BC)} = \frac{1}{\sqrt{2}} (|01\rangle \pm |10\rangle) \nonumber\\
|\phi^{\pm}\rangle_{AB(BC)} = \frac{1}{\sqrt{2}} (|00\rangle \pm |11\rangle)
\label{initial}
\end{eqnarray}
where, $\lbrace|0\rangle,|1\rangle\rbrace$ form eigenbasis of $\sigma_z$. Subsequent calculations are done with respect to the above specific initial state.

According to case-\ref{c1}, only Alice's particle undergoes damping and the state shared between Alice and Bob after environmental interaction becomes either,
\begin{equation}
\rho_{AB}' = \sum_{i=0}^1 (K_i \otimes \openone_2) ~ |\psi^{\pm}\rangle_{AB}\langle\psi^{\pm}| ~ (K_i^{\dagger} \otimes \openone_2)
\end{equation}
when the initial state between Alice and Bob is $|\psi^{\pm}\rangle_{AB}$ or,
\begin{equation}
\sigma_{AB}' = \sum_{i=0}^1 (K_i \otimes \openone_2) ~ |\phi^{\pm}\rangle_{AB}\langle\phi^{\pm}| ~ (K_i^{\dagger} \otimes \openone_2)
\end{equation}
when the initial state between Alice and Bob is $|\phi^{\pm}\rangle_{AB}$. Initial states are defined by Eq.(\ref{initial}) and Kraus operators $K_i$ corresponding to the damping parameter $p$ is given by Eq.(\ref{damp}). The joint state among Alice, Bob and Charlie thus becomes either $\rho_{AB}' \otimes |\psi^{\pm}\rangle_{BC}\langle\psi^{\pm}|$ ,or $\sigma_{AB}' \otimes |\phi^{\pm}\rangle_{BC}\langle\phi^{\pm}|$.

Similarly, for case-\ref{c2}, we consider that both Alice and Charlie are undergoing damping with
the  same damping parameter $p$. In this case, the post-interaction state between Bob and Charlie becomes either,
\begin{equation}
\rho_{BC}' = \sum_{i=0}^1 (\openone_2 \otimes K_i) ~ |\psi^{\pm}\rangle_{BC}\langle\psi^{\pm}| ~ (\openone_2 \otimes K_i^{\dagger})
\end{equation}
or,
\begin{equation}
\sigma_{BC}' = \sum_{i=0}^1 (\openone_2 \otimes K_i) ~ |\phi^{\pm}\rangle_{BC}\langle\phi^{\pm}| ~ (\openone_2 \otimes K_i^{\dagger})
\end{equation}
depending on the initial state shared between Bob and Charlie given by Eq.(\ref{initial}). Hence, the tripartite state shared among Alice, Bob and Charlie becomes either $\rho_{AB}' \otimes \rho_{BC}'$, or $\sigma_{AB}' \otimes \sigma_{BC}'$.

We now discuss quantum non-bilocal correlations generated by the above states in three given scenarios of Bell-state measurements, as mentioned above.

\subsection{Binary inputs and outputs}

The bilocal scenario where Bob performs two Bell-state measurements(BSM) and executes two possible outcomes per measurement and both Alice and Charlie perform dichotomic projective qubit measurements, gives rise to Eq.(\ref{2-2}) as a necessary condition for bilocality. We obtain the quantum non-bilocal correlations under this scenario resulting from the cases mentioned above.

\subsubsection{Case-\ref{c1}}

Here we  evaluate the left hand side of Eq.(\ref{2-2}) employing the joint state, $\rho_{AB}' \otimes |\psi^{\pm}\rangle_{BC}\langle\psi^{\pm}|$ ($\sigma_{AB}' \otimes |\phi^{\pm}\rangle_{BC}\langle\phi^{\pm}|$). The maximum of the quantity $B^{22}$ can be calculated following the method described in\cite{andreoli}. Using Eq.(\ref{andrmax}), we find that irrespective of the choice of the initial Bell state
\begin{equation}
B^{22}_{max} = \sqrt{2\sqrt{1-p}}
\label{Ac1}
\end{equation}
This is obtained from Eq.(\ref{andr}) by maximizing over all possible measurement settings by Alice, Bob and Charlie. The optimal choice of measurement settings are given by, $\vec{a}_0 = (-0.77,-0.64,0)$, $\vec{a}_1 = (-0.64,0.77,0)$, $\vec{b}_0^A = (0.09,-0.99,0)$, $\vec{b}_1^A = (-0.09,-0.99,0)$, $\vec{b}_0^C = (0,0,1)$, $\vec{b}_1^C= (1,0,0)$, $\vec{c}_0 = (\frac{1}{\sqrt{2}},0,\frac{1}{\sqrt{2}})$ and $\vec{c}_1 = (-\frac{1}{\sqrt{2}},0,\frac{1}{\sqrt{2}})$.

We observe that the  quantum mechanical (QM) violation of the upper bound of inequality(\ref{2-2}) occurs  when $p<\frac{3}{4}$. This indicates a range of decoherence parameter, i.e., $p\in[\frac{3}{4},1]$ for which the system will no longer be quantum non-bilocal.  The state without decoherence, i.e. $|\psi^{\pm}\rangle_{AB}\langle\psi^{\pm}| \otimes |\psi^{\pm}\rangle_{BC}\langle\psi^{\pm}|$ ($|\phi^{\pm}\rangle_{AB}\langle\phi^{\pm}| \otimes |\phi^{\pm}\rangle_{BC}\langle\phi^{\pm}|$) is of course the maximally non-bilocal quantum state. In Fig.2 $B^{22}_{max}$ is plotted w.r.t. the decoherence parameter $p$. 

\begin{figure}[!ht]
\resizebox{6cm}{4cm}{\includegraphics{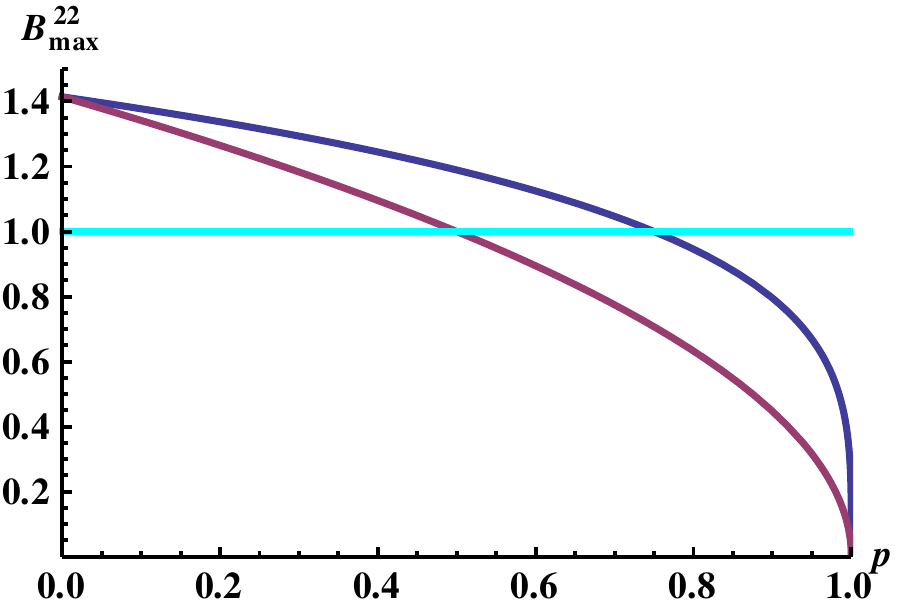}}
\caption{\footnotesize (Coloronline) $B^{22}_{max}$ is plotted w.r.t. decoherence parameter $p$ in the scenario of binary input and binary output(2-2). The upper  curve is for case-\ref{c1} when only Alice's part is affected by decoherence. The lower curve is for case-\ref{c2} when both Alice's and Charlie's subsystem interact with the environment. The straight line shows the  upper bound of bilocal correlations.}
\label{b22}
\end{figure}

\subsubsection{Case-\ref{c2}}

The state of the system after the interaction of the environment with both Alice's and Charlie's particle, i.e. $\rho_{AB}' \otimes \rho_{BC}'$ ($\sigma_{AB}' \otimes \sigma_{BC}'$) gives rise to the QM maximum of the quantity, $B^{22}$ according to Eq.(\ref{andrmax}) viz.
\begin{equation}
B^{22}_{max} = \sqrt{2(1-p)}
\label{Ac2}
\end{equation}
The above expression is obtained for the optimal choice of measurement settings for Alice, Bob and Charlie, given by, $\vec{a}_0 = (-0.78,-0.62,0)$, $\vec{a}_1 = (-0.62,0.78,0)$, $\vec{b}_0^A = (0.99,0.12,0)$, $\vec{b}_1^A = (-0.12,-0.99,0)$, $\vec{b}_0^C = (-0.85,0.52,0)$, $\vec{b}_1^C= (-0.52,-0.85,0)$, $\vec{c}_0 = (-0.97,-0.24,0)$ and $\vec{c}_1 = (-0.24,0.97,0)$. QM maximum of $B^{22}$ violates inequality(\ref{2-2}) for $p<\frac{1}{2}$. Thus, in the region of $p\in[\frac{1}{2},\frac{3}{4})$, the quantum non-bilocal correlation which survives under single environmental interaction with Alice,  disappears under the  environmental interaction of both Alice and Charlie.

Fig.\ref{b22} illustrates that as number of interaction with the environment increases, the amount of quantum non-bilocality reduces in the system. It may be noted that unlike the case of teleportation fidelity\cite{except2,except3}, quantum non-bilocality resembles the behaviour of the  lower bound of quantum secret key rate in one-sided device independent quantum key distribution(1s-DIQKD)\cite{weak3}, where classical correlation plays no role in improving the relevant measure under decoherence.

\subsection{One input and four outputs for Bob}

The bilocal correlation where Bob performs one BSM having four distinguishable outputs and Alice and Charlie perform binary qubit measurements having two outcomes per measurement, is displayed through the bilocality inequality(\ref{1-4}). In this scenario, we discuss quantum non-bilocal correlations arising from the afore-mentioned cases of decohering systems.

\subsubsection{Case-\ref{c1}}

Considering the state, $\rho_{AB}' \otimes |\psi^{\pm}\rangle_{BC}\langle\psi^{\pm}|$ ($\sigma_{AB}' \otimes |\phi^{\pm}\rangle_{BC}\langle\phi^{\pm}|$), we find the QM maximum of the left hand side of Eq.(\ref{1-4}), regardless of the choice of the initial Bell state as
\begin{equation}
B^{14}_{max} = \sqrt{(1+\sqrt{1-p})\sqrt{1-p}}
\label{Bc1}
\end{equation}
This is accomplished for the choice of observables $\lbrace A_0 \equiv \frac{1}{\sqrt{1+\sqrt{1-p}}} \sigma_x + \frac{\sqrt{\sqrt{1-p}}}{\sqrt{1+\sqrt{1-p}}} \sigma_z, A_1 \equiv - \frac{1}{\sqrt{1+\sqrt{1-p}}} \sigma_x + \frac{\sqrt{\sqrt{1-p}}}{\sqrt{1+\sqrt{1-p}}} \sigma_z \rbrace$ at Alice's side and the choice of observables $\lbrace C_0 \equiv \frac{1}{\sqrt{1+\sqrt{1-p}}} \sigma_x + \frac{\sqrt{\sqrt{1-p}}}{\sqrt{1+\sqrt{1-p}}} \sigma_z, C_1 \equiv - \frac{1}{\sqrt{1+\sqrt{1-p}}} \sigma_x + \frac{\sqrt{\sqrt{1-p}}}{\sqrt{1+\sqrt{1-p}}} \sigma_z \rbrace$ corresponding to Charlie's side. It can be checked that the correlation $B^{14}_{max}$ is non-bilocal for $p<\frac{\sqrt{5}-1}{2}\simeq 0.62$. 

\begin{figure}[!ht]
\resizebox{6cm}{4cm}{\includegraphics{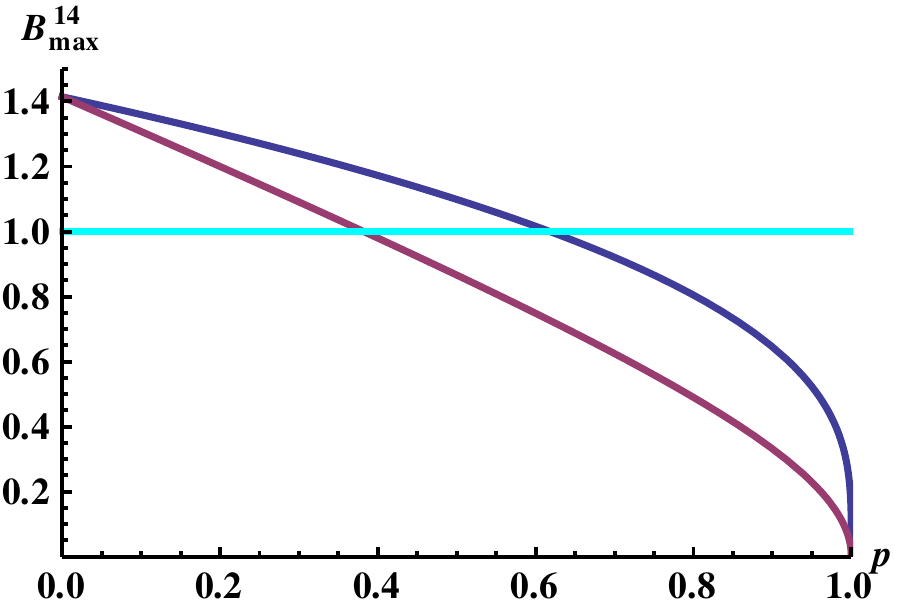}}
\caption{\footnotesize (Coloronline) $B^{14}_{max}$ is plotted against decoherence parameter $p$. The
upper and lower curves represent Eq.(\ref{Bc1}) of case-\ref{c1} and Eq.(\ref{Bc2}) of case-\ref{c2}, respectively. The straight line shows the upper bound of bilocal correlations.}
\label{b14}
\end{figure}

\subsubsection{Case-\ref{c2}}

For the state given by, $\rho_{AB}' \otimes \rho_{BC}'$ ($\sigma_{AB}' \otimes \sigma_{BC}'$), the QM maximum of the quantity, $B^{14}$ appearing in Eq.(\ref{1-4}) is given by
\begin{equation}
B^{14}_{max} = \sqrt{(1-p)(2-p)}
\label{Bc2}
\end{equation}
This occurs for the the choice of observables $\lbrace A_0 \equiv \frac{1}{\sqrt{2-p}} \sigma_x + \sqrt{\frac{1-p}{2-p}} \sigma_z, A_1 \equiv - \frac{1}{\sqrt{2-p}} \sigma_x + \sqrt{\frac{1-p}{2-p}} \sigma_z \rbrace$ at Alice's side and $\lbrace C_0 \equiv \frac{1}{\sqrt{2-p}} \sigma_x + \sqrt{\frac{1-p}{2-p}} \sigma_z, C_1 \equiv - \frac{1}{\sqrt{2-p}} \sigma_x + \sqrt{\frac{1-p}{2-p}} \sigma_z \rbrace$ at Charlie's side.
It is observed that, $B^{14}_{max}$ violates inequality(\ref{1-4}) for $p< \frac{3-\sqrt{5}}{2} \simeq 0.38$. Here too, the quantum non-bilocal correlation lowers with the presence of more interaction with the environment, as depicted in Fig.\ref{b14}.

\subsection{Scenario of one input and three outputs for Bob}

Now we consider the bilocal scenario where Bob has one complete BSM as input and produces three outputs, whereas Alice and Charlie have binary inputs and binary outputs. The corresponding bilocality inequality is given by Eq.(\ref{1-3}). Similar to the previous scenarios, here we analyze the  two relevant cases in the following.

\subsubsection{Case-\ref{c1}}

If we employ the state $\rho_{AB}' \otimes |\psi^{\pm}\rangle_{BC}\langle\psi^{\pm}|$ ($\sigma_{AB}' \otimes |\phi^{\pm}\rangle_{BC}\langle\phi^{\pm}|$) to compute the QM maximum of the quantity $B^{13}$ given by Eq.(\ref{1-3}), it turns out that
\begin{equation}
B^{13}_{max} = \sqrt{\frac{\sqrt{1-p}(1+2\sqrt{1-p})}{2}}
\label{Cc1}
\end{equation}
This is achieved by maximizing over all possible measurement settings of Alice and Charlie. The corresponding measurement settings are given by $\lbrace A_0 \equiv - \frac{1}{\sqrt{1+2\sqrt{1-p}}} \sigma_x + \sqrt{\frac{2\sqrt{1-p}}{1+2\sqrt{1-p}}} \sigma_z, A_1 \equiv \frac{1}{\sqrt{1+2\sqrt{1-p}}} \sigma_x + \sqrt{\frac{2\sqrt{1-p}}{1+2\sqrt{1-p}}} \sigma_z \rbrace$ for Alice, and $\lbrace C_0 \equiv - \frac{1}{\sqrt{1+2\sqrt{1-p}}} \sigma_x + \sqrt{\frac{2\sqrt{1-p}}{1+2\sqrt{1-p}}} \sigma_z, C_1 \equiv \frac{1}{\sqrt{1+2\sqrt{1-p}}} \sigma_x + \sqrt{\frac{2\sqrt{1-p}}{1+2\sqrt{1-p}}} \sigma_z \rbrace$ for Charlie.
Therefore, $B^{13}_{max}$ violates inequality(\ref{1-3}) when $p< \frac{\sqrt{17}-1}{8} \simeq 0.39$.

\begin{figure}[!ht]
\resizebox{6cm}{4cm}{\includegraphics{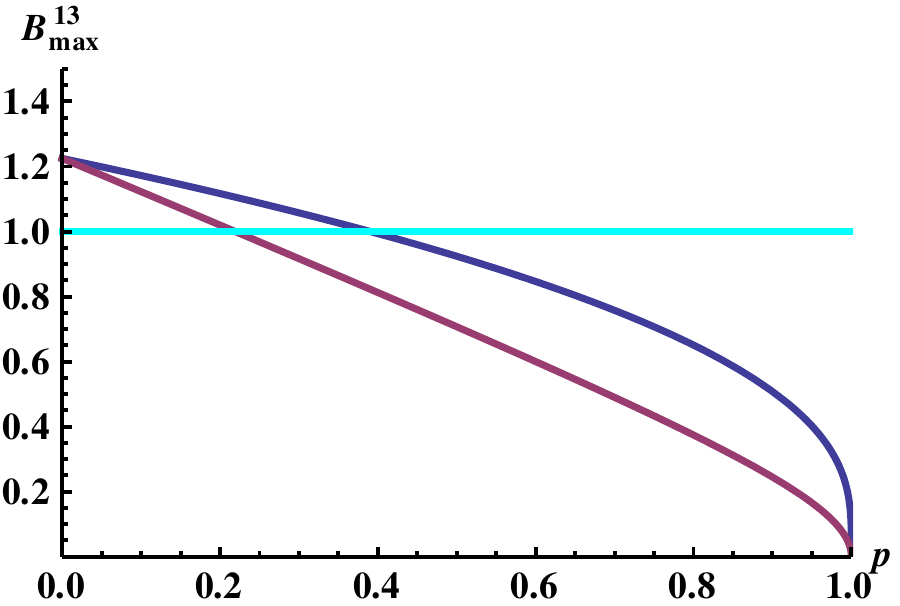}}
\caption{\footnotesize $B^{13}_{max}$ is plotted w.r.t. the decoherence parameter $p$ for any of the initial Bell states. The upper curve representing case-\ref{c1} and the lower curve representing case-\ref{c2} signify Eq.(\ref{Cc1}) and Eq.(\ref{Cc2}) respectively. The straight line is the upper bound of bilocal correlations.}
\label{b13}
\end{figure}

\subsubsection{Case-\ref{c2}}

Now using the state given by $\rho_{AB}' \otimes \rho_{BC}'$ ($\sigma_{AB}' \otimes \sigma_{BC}'$), we find that $B^{13}$ reaches its QM maximum given by
\begin{equation}
B^{13}_{max} = \sqrt{\frac{(1-p)(3-2p)}{2}}
\label{Cc2}
\end{equation}
when Alice chooses the observables, $\lbrace A_0 \equiv - \frac{1}{\sqrt{3-2p}} \sigma_x + \sqrt{\frac{2(1-p)}{3-2p}} \sigma_z, A_1 \equiv \frac{1}{\sqrt{3-2p}} \sigma_x + \sqrt{\frac{2(1-p)}{3-2p}} \sigma_z \rbrace$, and Charlie chooses  $\lbrace C_0 \equiv \frac{1}{\sqrt{3-2p}} \sigma_x + \sqrt{\frac{2(1-p)}{3-2p}} \sigma_z, C_1 \equiv - \frac{1}{\sqrt{3-2p}} \sigma_x + \sqrt{\frac{2(1-p)}{3-2p}} \sigma_z \rbrace$.
Quantum non-bilocal correlation is exhibited for the domain of the decoherence parameter, $p\in[0,\frac{5-\sqrt{17}}{4} \simeq 0.22)$. We see that the range of violation w.r.t. $p$ narrows for this
case compared to the previous scenarios with other forms of BSM, as  is shown in Fig.\ref{b13}.

\section{Shielding of non-bilocal correlations using weak measurement} \label{5}

There are several examples of application of the  weak measurement technique  to protect quantum correlations  in the literature\cite{weak1,weak2,weak3,wktch4,wktch5,wktch6}. This procedure involves the action of weak measurement before the effect of decoherence, and the performance of reverse weak measurement after the state has undergone decoherence. We have already discussed two possible cases, where only Alice's particle undergoes decoherence by means of ADC and where both Alice's and Charlie's particles are passed through ADC. Corresponding to these cases, there appears two possible circumstances of applying the technique of weak measurement, i.e.
\begin{enumerate}
\item \label{d1} Bob performs weak measurement before sending the particle to Alice through ADC, and after receiving the subsystem, Alice performs reverse weak measurement on it.
\item \label{d2} Bob applies weak measurement on both of the subsystems which are to be transmitted individually to Alice and Charlie across ADC. When Alice and Charlie receive their subsequent particles, they both perform reverse weak measurement separately.
\end{enumerate}

According to case-\ref{d1}, Bob performs weak measurement on the Alice's part of the initial joint state specified in Eq.(\ref{initial}) in a tripartite network. When Alice's subsystem is not detected with probability $w$, the joint state between Alice and Bob becomes
\begin{equation}
\rho_{AB}^{w} = (W_1 \otimes \openone_2) ~ |\psi^{\pm}\rangle_{AB}\langle\psi^{\pm}| ~ (W_1^{\dagger} \otimes \openone_2)
\label{w1}
\end{equation}
when the initial state is $|\psi^{\pm}\rangle_{AB}$, or
\begin{equation}
\sigma_{AB}^{w} = (W_1 \otimes \openone_2) ~ |\phi^{\pm}\rangle_{AB}\langle\phi^{\pm}| ~ (W_1^{\dagger} \otimes \openone_2)
\label{w2}
\end{equation}
corresponding to the initial state $|\phi^{\pm}\rangle_{AB}$. The operator $W_1$ is given by Eq.(\ref{weak}). Hence, the unnormalised joint state between Alice, Bob and Charlie becomes $\rho_{AB}^{w} \otimes |\psi^{\pm}\rangle_{BC}\langle\psi^{\pm}|$ ($\sigma_{AB}^{w} \otimes |\phi^{\pm}\rangle_{BC}\langle\phi^{\pm}|$). The success probability associated with the post-selection is
\begin{align}
P_{succ}^{w} = & \operatorname{Tr}[\rho_{AB}^{w} \otimes |\psi^{\pm}\rangle_{BC}\langle\psi^{\pm}|] \nonumber\\
= & \operatorname{Tr}[\sigma_{AB}^{w} \otimes |\phi^{\pm}\rangle_{BC}\langle\phi^{\pm}|] \nonumber\\
= & 1-\frac{w}{2}
\end{align}

After the effect of ADC, Alice and Bob share the  state 
\begin{equation}
\rho_{AB}^{d} = \sum_{i=0}^1 (K_i \otimes \openone_2) ~ \rho_{AB}^{w} ~ (K_i^{\dagger} \otimes \openone_2)
\end{equation}
when the initial state is $|\psi^{\pm}\rangle_{AB}$, or
\begin{equation}
\sigma_{AB}^{d} = \sum_{i=0}^1 (K_i \otimes \openone_2) ~ \sigma_{AB}^{w} ~ (K_i^{\dagger} \otimes \openone_2)
\end{equation}
when the initial state is $|\phi^{\pm}\rangle_{AB}$. $\lbrace K_i \rbrace_{i=0,1}$ are the Kraus operators represented by Eq.(\ref{damp}). This results into the tripartite joint state $\rho_{AB}^{d} \otimes |\psi^{\pm}\rangle_{BC}\langle\psi^{\pm}|$ ($\sigma_{AB}^{d} \otimes |\phi^{\pm}\rangle_{BC}\langle\phi^{\pm}|$).
Then Alice performs another (reverse) of weak measurement to end up with the following state shared with Bob:
\begin{equation}
\rho_{AB}^{r} = (R_1 \otimes \openone_2) ~ \rho_{AB}^{d} ~ (R_1^{\dagger} \otimes \openone_2)
\label{r1}
\end{equation}
corresponding to the initial state $|\psi^{\pm}\rangle_{AB}$, or
\begin{equation}
\sigma_{AB}^{r} = (R_1 \otimes \openone_2) ~ \sigma_{AB}^{d} ~ (R_1^{\dagger} \otimes \openone_2)
\label{r2}
\end{equation}
corresponding to the initial state $|\phi^{\pm}\rangle_{AB}$.  The Kraus operator $R_1$ is defined in Eq.(\ref{reverse}). This leads to the joint state $\rho_{AB}^{r} \otimes |\psi^{\pm}\rangle_{BC}\langle\psi^{\pm}|$ ($\sigma_{AB}^{r} \otimes |\phi^{\pm}\rangle_{BC}\langle\phi^{\pm}|$) among the three parties. Now we normalise the above state by taking into account the success probability of obtaining $\rho_{AB}^{r}$ ($\sigma_{AB}^{r}$), i.e.,
\begin{align}
P_{succ}^{r} = & \operatorname{Tr}[\rho_{AB}^{r} \otimes |\psi^{\pm}\rangle_{BC}\langle\psi^{\pm}|] \nonumber\\
= & \operatorname{Tr}[\sigma_{AB}^{r} \otimes |\phi^{\pm}\rangle_{BC}\langle\phi^{\pm}|]
\end{align}

As per case-\ref{d2}, Bob performs weak measurement not only on Alice's subsystem, but also on Charlie's subsystem with strength $w$ which we consider to be the same for both the subsystems for
simplicity. So, the bipartite state of Bob and Charlie becomes
\begin{equation}
\rho_{BC}^{w} = (\openone_2 \otimes W_1) ~ |\psi^{\pm}\rangle_{BC}\langle\psi^{\pm}| ~ (\openone_2 \otimes W_1^{\dagger})
\end{equation}
when the initial state is $|\psi^{\pm}\rangle_{BC}$, or
\begin{equation}
\sigma_{BC}^{w} = (\openone_2 \otimes W_1) ~ |\phi^{\pm}\rangle_{BC}\langle\phi^{\pm}| ~ (\openone_2 \otimes W_1^{\dagger})
\end{equation}
when the initial state is $|\phi^{\pm}\rangle_{BC}$.
The  form of $W_1$ is given by Eq.(\ref{weak}). The full tripartite  state takes the form $\rho_{AB}^{w}\otimes\rho_{BC}^{w}$ ($\sigma_{AB}^{w}\otimes\sigma_{BC}^{w}$), where $\rho_{AB}^{w}$($\sigma_{AB}^{w}$) is defined by Eq.(\ref{w1}) (Eq.(\ref{w2})). The success probability taking into account the  post-selections done on the system becomes
\begin{align}
P_{succ}^{ww} =  & \operatorname{Tr}[\rho_{AB}^{w} \otimes \rho_{BC}^{w}] = \operatorname{Tr}[\sigma_{AB}^{w} \otimes \sigma_{BC}^{w}] \nonumber\\
= & (1-\frac{w}{2})^2
\end{align}

Now, a couple of independent decoherence channels (ADC) are applied on both of the particles, on effect of which the joint state between Alice, Bob and Charlie transforms as
\begin{align}
\rho_{AB}^{d} \otimes \rho_{BC}^{d} = \sum_{i,j=0}^{1} & (K_i \otimes \openone_2 \otimes \openone_2 \otimes K_j) ~ (\rho_{AB}^{w} \otimes \rho_{BC}^{w}) \nonumber\\
& (K_i^{\dagger} \otimes \openone_2 \otimes \openone_2 \otimes K_j^{\dagger})
\end{align}
or
\begin{align}
\sigma_{AB}^{d} \otimes \sigma_{BC}^{d} = \sum_{i,j=0}^{1} & (K_i \otimes \openone_2 \otimes \openone_2 \otimes K_j) ~ (\sigma_{AB}^{w} \otimes \sigma_{BC}^{w}) \nonumber\\
& (K_i^{\dagger} \otimes \openone_2 \otimes \openone_2 \otimes K_j^{\dagger})
\end{align}
where, $\rho_{BC}^{d} = \sum_{j=0}^{1} (\openone_2 \otimes K_j) \rho_{BC}^{w} (\openone_2 \otimes K_j^{\dagger})$ and $\sigma_{BC}^{d} = \sum_{j=0}^{1} (\openone_2 \otimes K_j) \sigma_{BC}^{w} (\openone_2 \otimes K_j^{\dagger})$, depending upon the initial states given by Eq.(\ref{initial}). Here also for simplicity we choose the same parameter of decoherence $p$ for both Alice and Charlie.

Next, Alice and Charlie make reverse weak measurements on their respective particles with the same strength $r$.  After completion of the above procedures (resulting in entanglement swapping from the Bob-Alice and the Bob-Charlie pairs to the Alice-Charlie pair), the final Bob-Charlie state becomes
\begin{equation}
\rho_{BC}^{r} = (\openone_2 \otimes R_1) ~ \rho_{BC}^{d} ~ (\openone_2 \otimes R_1^{\dagger})
\end{equation}
with reference to the initial state $|\psi^{\pm}\rangle_{BC}$, or
\begin{equation}
\sigma_{BC}^{r} = (\openone_2 \otimes R_1) ~ \sigma_{BC}^{d} ~ (\openone_2 \otimes R_1^{\dagger})
\end{equation}
with reference to the initial state $|\phi^{\pm}\rangle_{BC}$  ($R_1$ is given by Eq.(\ref{reverse})), where state normalization is incorporated  by calculating the success probability of the  post-selection procedures in the reverse weak measurement processes, given by
\begin{equation}
P_{succ}^{rr} = \operatorname{Tr}[\rho_{AB}^{r} \otimes \rho_{BC}^{r}] = \operatorname{Tr}[\sigma_{AB}^{r} \otimes \sigma_{BC}^{r}]
\end{equation}
We are now in a position to discuss  bilocality of the final state in context of the above two cases in three different scenarios of Bell-state measurements.

\subsection{Binary inputs and outputs}

In this scenario, Eq.(\ref{2-2}) is the necessary condition for bilocality. There arises the following two cases involving the use of the weak measurement technique.

\subsubsection{Case-\ref{d1}}

With the help of Eq.(\ref{andr}) and the joint state, $\rho_{AB}^{r} \otimes |\psi^{\pm}\rangle_{BC}\langle\psi^{\pm}|$ ($\sigma_{AB}^{r} \otimes |\phi^{\pm}\rangle_{BC}\langle\phi^{\pm}|$) we first calculate the quantity $B^{22}$ keeping the measurement settings for  Alice, Bob and Charlie similar to the case when the weak measurement technique is not used. The violation of Eq.(\ref{2-2}) is maximized by suitably choosing the reverse weak parameter $r$ by Alice. Hence,
\begin{equation}
B^{22} = \frac{\sqrt{2}}{(1+p-pw)^{\frac{1}{4}}}
\label{Ad1}
\end{equation}
is achieved for the optimal strength of the reverse weak measurement given by $r_{opt} = \frac{2p+w-2pw}{1+p-pw}$. This makes the success probability $P_{succ}^r = (1-p)(1-w)$.

\begin{figure}
\begin{subfigure}{4.3cm}
  \centering
  \includegraphics[width=4.3cm]{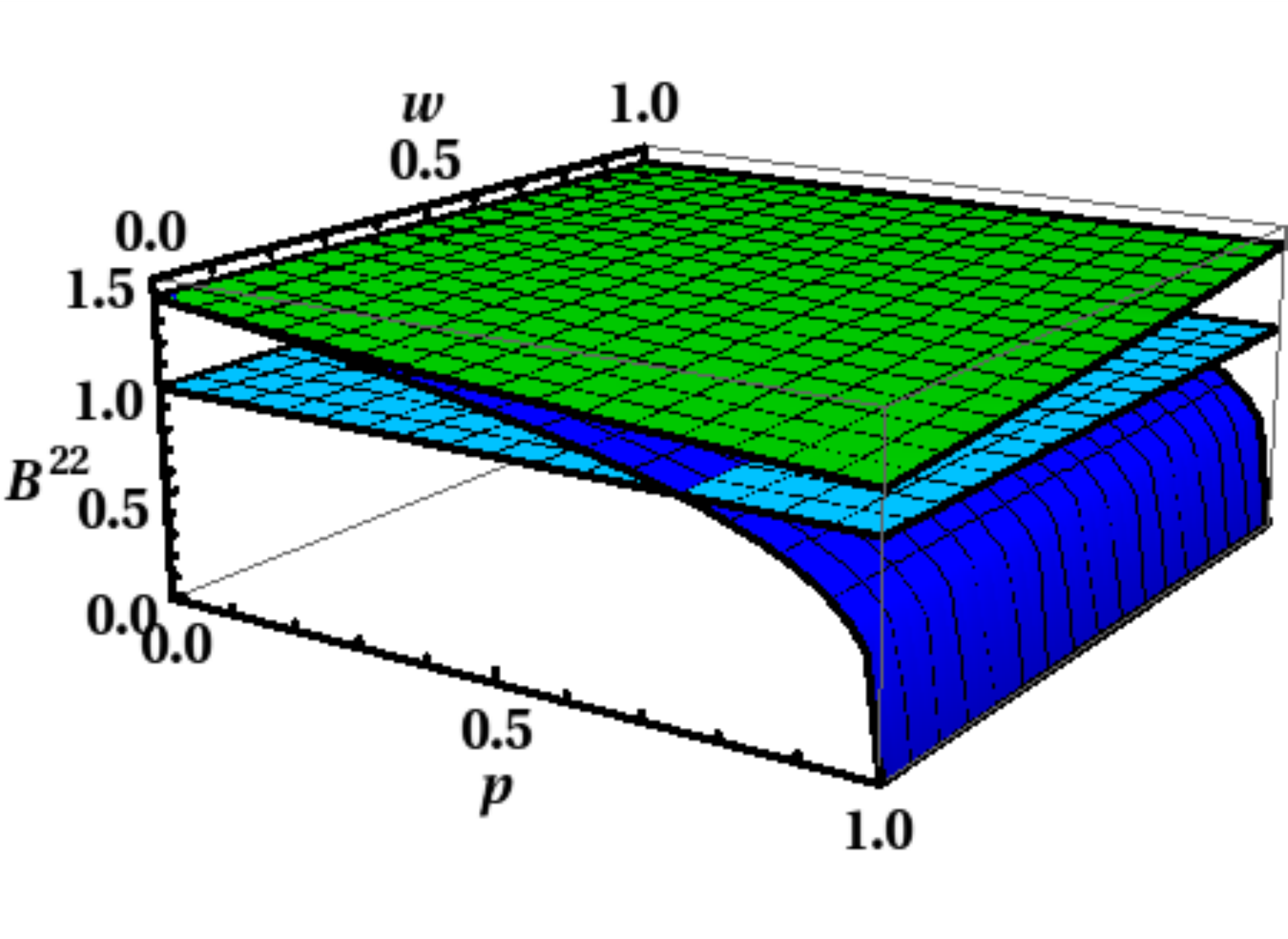}
  \caption{\footnotesize case-\ref{d1}}
  \label{1w22}
\end{subfigure}%
\begin{subfigure}{4.3cm}
  \centering
  \includegraphics[width=4.3cm]{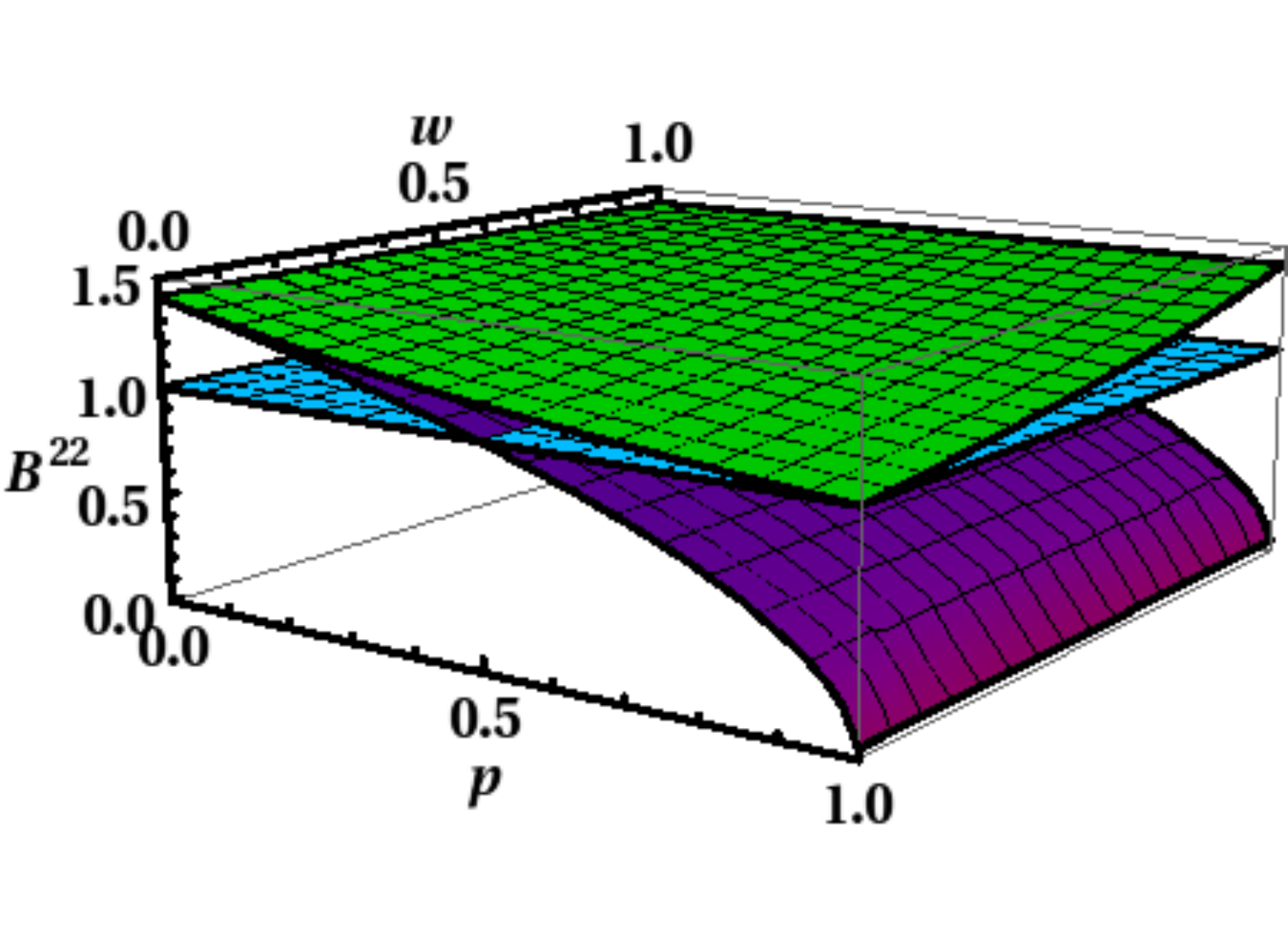}
  \caption{\footnotesize case-\ref{d2}}
  \label{2w22}
\end{subfigure}
\caption{\footnotesize (Coloronline) The left hand side of bilocality inequality $B^{22}$ is plotted w.r.t. the damping parameter $p$(x-axis) and the weak measurement parameter $w$(y-axis). The upper  surfaces in (\ref{1w22}) and (\ref{2w22}) denote Eq.(\ref{Ad1}) and Eq.(\ref{Ad2}) respectively, when the technique of weak measurement is used by optimizing over the parameter $r$. The lower  surfaces 
correspond to the analogous cases without using weak measurement. The flat surface shows the upper bound of bilocal correlations.}
\label{w22}
\end{figure}

If we compare the left hand side of Eq.(\ref{2-2}), i.e., $B^{22}_{max}$ given by Eq.(\ref{Ac1}) when the technique of weak measurement is not incorporated, with the quantity $B^{22}$ given by Eq.(\ref{Ad1}) when the technique of weak measurement is performed,  it is observed from Fig.\ref{1w22} that the latter quantity is quantum non-bilocal for the entire range of the parameters $p$ and $w$. Therefore, it is clear that like certain other correlations\cite{weak1,weak2,weak3}, the quantum non-bilocal correlation is preserved under the effect of decoherence (ADC) if the technique of weak measurement is applied.
\subsubsection{Case-\ref{d2}}

Using Eq.(\ref{andr}), the left hand side of Eq.(\ref{2-2}) for the whole system, $\rho_{AB}^{r} \otimes \rho_{BC}^{r}$ ($\sigma_{AB}^{r} \otimes \sigma_{BC}^{r}$) is computed keeping the measurement settings fixed w.r.t. the situation where weak measurement technique is not engaged.  Then, maximizing over the reverse weak parameter ($r$) of Alice and Charlie, one gets
\begin{equation}
B^{22} = \sqrt{\frac{2}{1+p(1-w)}}
\label{Ad2}
\end{equation}
This occurs for the optimal reverse weak parameter given by $r_{opt} = \frac{2p+w-2pw}{1+p-pw}$. Corresponding to this, the success probability of achieving $\rho_{AB}^{r} \otimes \rho_{BC}^{r}$ ($\sigma_{AB}^{r} \otimes \sigma_{BC}^{r}$)  becomes, $P_{succ}^{rr} = (1-p)^2 (1-w)^2$.
It can be easily checked that the expression in Eq.(\ref{Ad2}) is always greater than or equal to $1$. Hence, in comparison to Eq.(\ref{Ac2}), it can be seen that the weak measurement technique provides improvement in the manifestation of quantum non-bilocality from the case without the use of it. This is displayed in Fig.\ref{2w22}.

\subsection{One input and four outputs for Bob}

The bilocality in this scenario is captured by the necessary condition given by Eq.(\ref{1-4}). The application of the weak measurement technique is described in the following cases.

\subsubsection{Case-\ref{d1}}

The tripartite joint state $\rho_{AB}^{r} \otimes |\psi^{\pm}\rangle_{BC}\langle\psi^{\pm}|$ ($\sigma_{AB}^{r} \otimes |\phi^{\pm}\rangle_{BC}\langle\phi^{\pm}|$) contingent upon the measurement settings chosen by Alice and Charlie to be the same as that of the case when the weak measurement technique is not employed, leads to
\begin{align}
B^{14} = & \sqrt{\frac{\sqrt{1-p}}{1+\sqrt{1-p}}} ~ \frac{1}{\sqrt{2-[1+p(1-w)] r -w}} \nonumber\\
& \times \Big(\sqrt{[1-p(1-w)](2-r)-w} \nonumber\\
& + \sqrt{2\sqrt{(1-w)(1-r)}}\Big)
\end{align}

\begin{figure}
\begin{subfigure}{4.3cm}
  \centering
  \includegraphics[width=4.3cm]{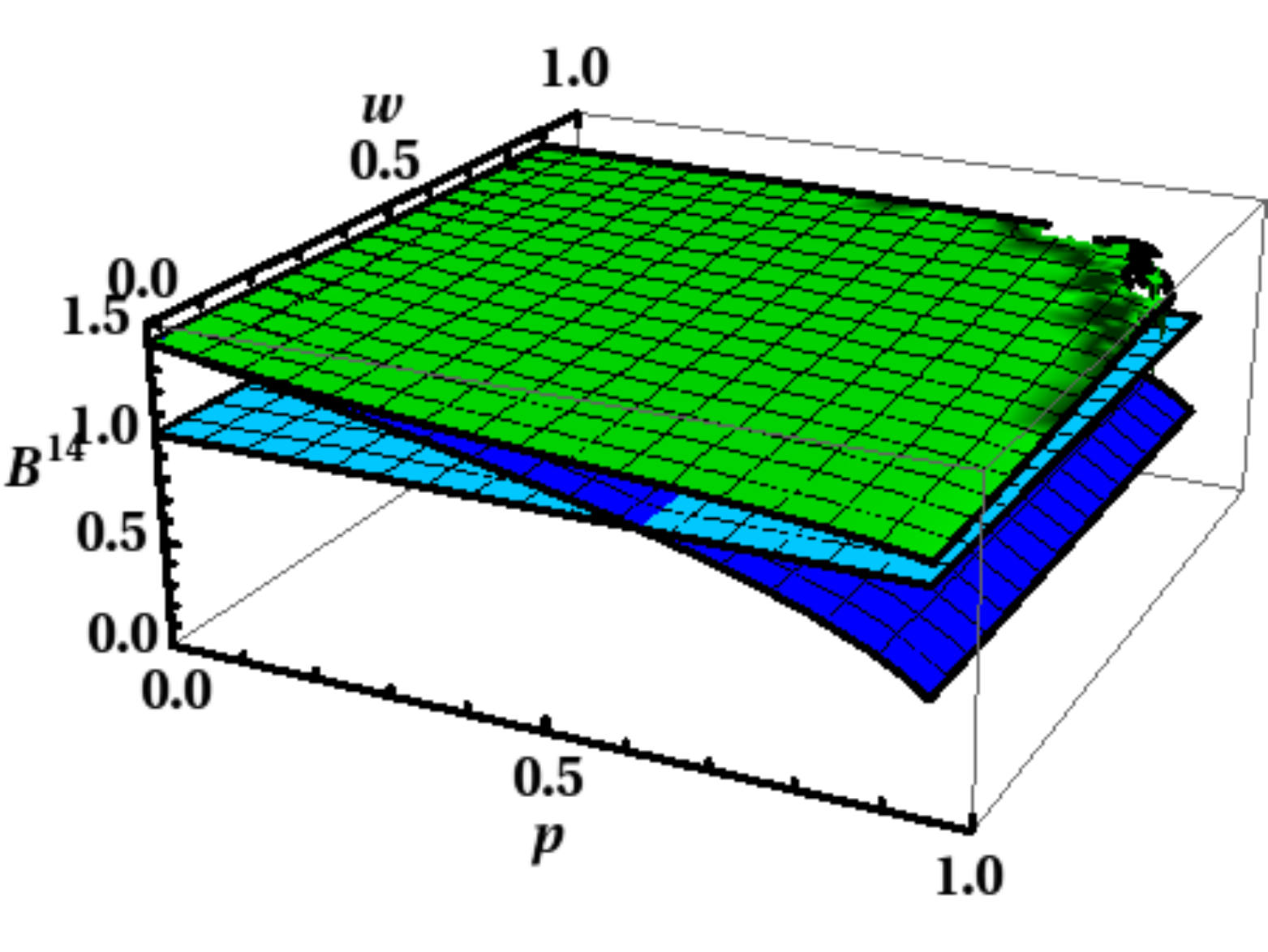}
  \caption{\footnotesize case-\ref{d1}}
  \label{1w14}
\end{subfigure}%
\begin{subfigure}{4.3cm}
  \centering
  \includegraphics[width=4.3cm]{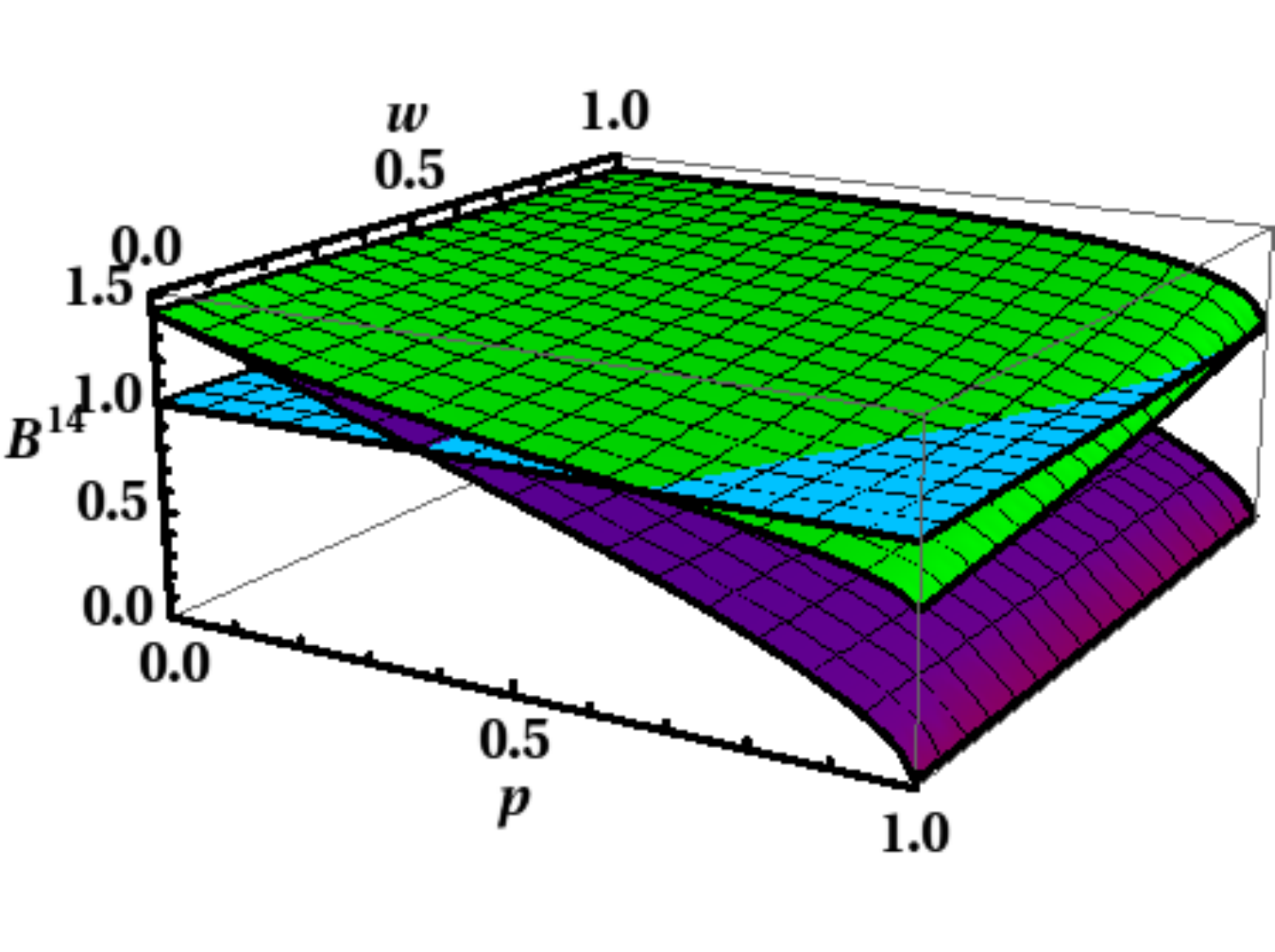}
  \caption{\footnotesize case-\ref{d2}}
  \label{2w14}
\end{subfigure}
\caption{\footnotesize (Coloronline) $B^{14}$ is plotted w.r.t. the damping parameter $p$ (x-axis) and the weak measurement parameter $w$ (y-axis). The upper  surfaces in (\ref{1w14}) and (\ref{2w14}) are obtained using the technique of weak measurement and the lower  surfaces are obtained without applying weak measurement. The flat surface shows the upper bound of bilocal correlations.}
\label{w14}
\end{figure}

The above expression is maximized w.r.t. the reverse weak parameter $r$ chosen by Alice. The
results are displayed  in the  Fig.\ref{1w14}, showing that, in compared to Eq.(\ref{Bc1}), the weak measurement technique preserves non-bilocal correlation throughout the range of $p$ and $w$.

\subsubsection{Case-\ref{d2}}

If the measurement settings for Alice and Charlie are kept unaltered, then using the state $\rho_{AB}^{r} \otimes \rho_{BC}^{r}$ ($\sigma_{AB}^{r} \otimes \sigma_{BC}^{r}$), the left hand side of Eq.(\ref{1-4}) becomes
\begin{align}
B^{14} = & \sqrt{\frac{1-p}{2-p}} ~ \frac{1}{2-[1+p(1-w)]r-w} \nonumber\\
& \times \Big([1-p(1-w)](2-r)-w \nonumber\\
& +2\sqrt{(1-w)(1-r)}\Big)
\end{align}

The above expression is maximized over the strength of the reverse weak measurement($r$) done by Alice and Charlie. Thus, the optimum strength turns out to be $r_{opt}= \frac{1}{(1+p-pw)^2} \Big( w + p(1-w) [(2+w) + 2x(1-w) +2p^2(2-p)(1-w)^2 + 2pw(1-w)]\Big)$, where $x=(1-p)\sqrt{(1-p)+p(1-p)(1-w)+p^2 (1-p)^2 (1-w)^2}$. Using $r=r_{opt}$, one can plot $B^{14}$ w.r.t $p$ and $w$  and compare it with Eq.(\ref{Bc2}). The illustration given by Fig.\ref{2w14} showing that the weak measurement technique is  able to slow down the effect of decoherence and thereby, exhibit improved quantum non-bilocal correlation for a considerable range of $p$ and $w$.

\subsection{One input and three outputs for Bob}

We now come discuss the scenario where bilocality is encapsulated by the inequality(\ref{1-3}) in the context of the weak measurement technique in the following cases. 

\subsubsection{Case-\ref{d1}}

Similar to the previous scenarios, here we compute the left hand side of Eq.(\ref{1-3}) with the help of the joint state $\rho_{AB}^{r} \otimes |\psi^{\pm}\rangle_{BC}\langle\psi^{\pm}|$ ($\sigma_{AB}^{r} \otimes |\phi^{\pm}\rangle_{BC}\langle\phi^{\pm}|$) and the measurement settings which are optimal for the case  without application of weak measurement. Hence,
\begin{align}
B^{13} = & \sqrt{\frac{\sqrt{1-p}}{1+2\sqrt{1-p}}} ~ \frac{1}{\sqrt{2-[1+p(1-w)] r -w}} \nonumber\\
& \times \Big(\sqrt{2[1-p(1-w)](2-r)-2w} \nonumber\\
& + \sqrt{\sqrt{(1-w)(1-r)}}\Big)
\end{align}

\begin{figure}
\begin{subfigure}{4.3cm}
  \centering
  \includegraphics[width=4.3cm]{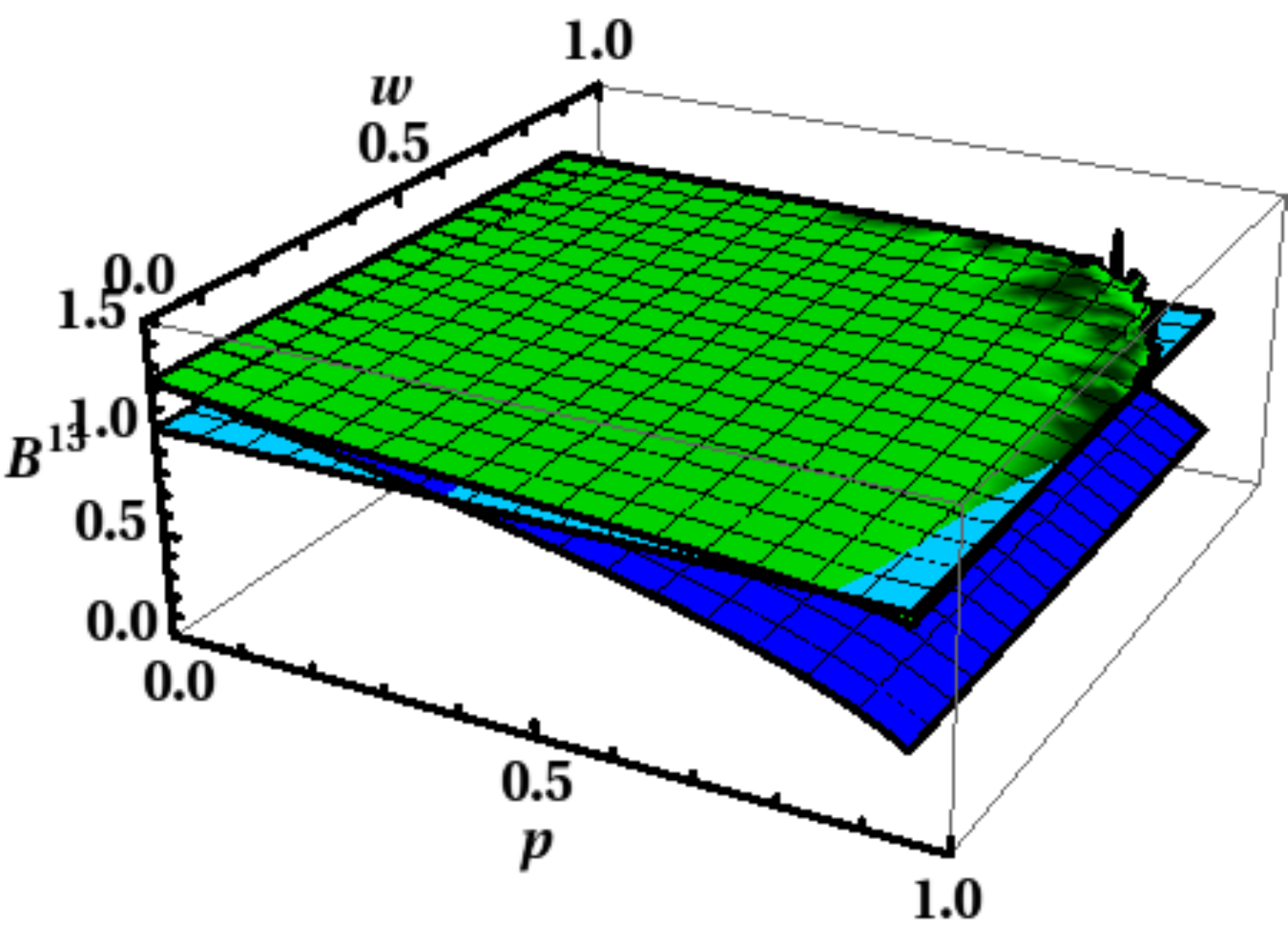}
  \caption{\footnotesize case-\ref{d1}}
  \label{1w13}
\end{subfigure}%
\begin{subfigure}{4.3cm}
  \centering
  \includegraphics[width=4.3cm]{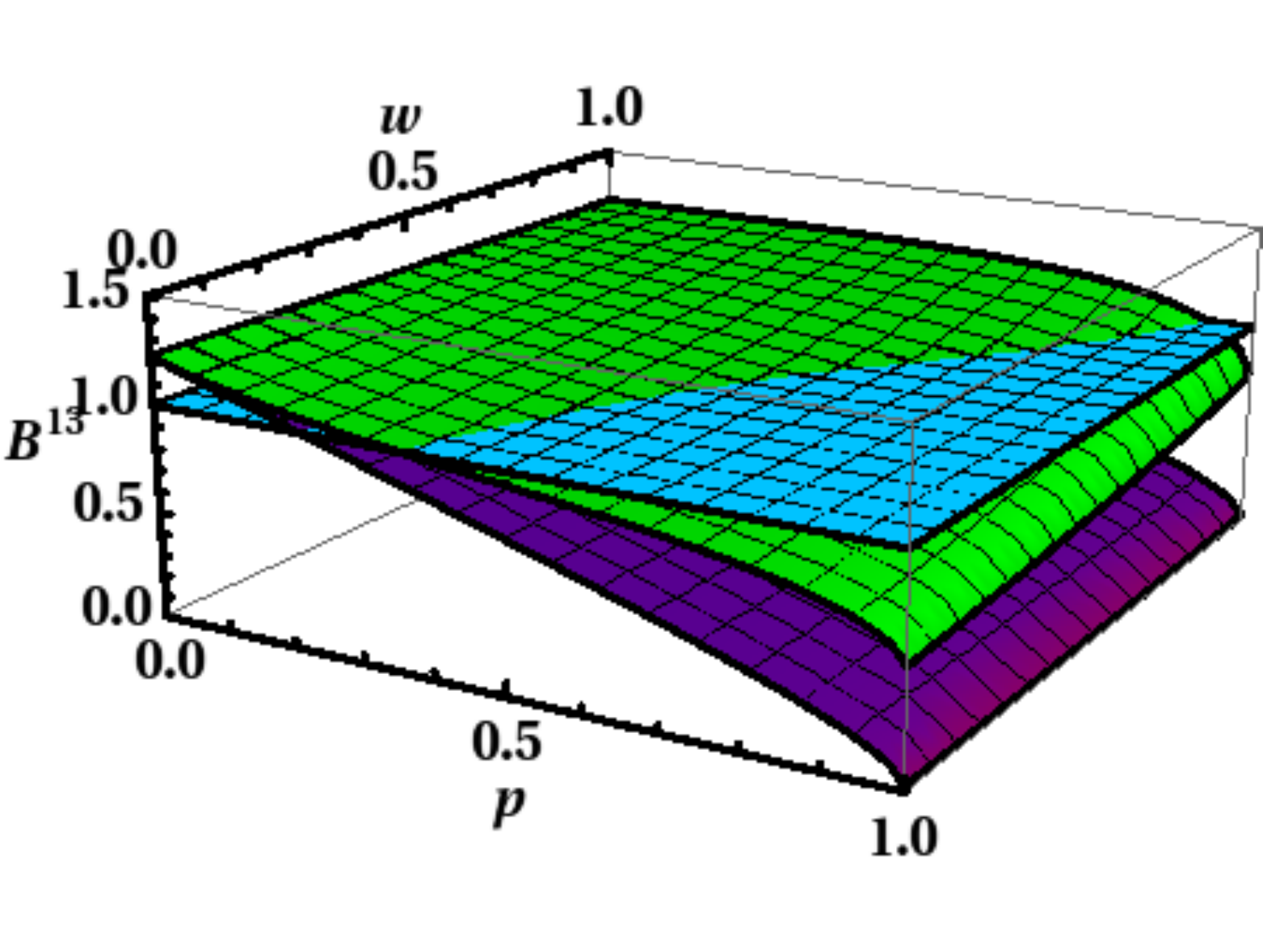}
  \caption{\footnotesize case-\ref{d2}}
  \label{2w13}
\end{subfigure}
\caption{\footnotesize (coloronline) $B^{13}$ is plotted w.r.t. the damping parameter $p$ (x-axis) and the weak parameter $w$ (y-axis) where the upper  surfaces are obtained by applying the technique of weak measurement and the lower  surfaces in (\ref{1w13}) and (\ref{2w13}) correspond to Eq.(\ref{Cc1}) and Eq.(\ref{Cc2}), respectively. The flat surface  shows the  upper bound of bilocal correlations.}
\label{w13}
\end{figure}

After maximizing the above function w.r.t. the reverse weak strength $r$ for Alice, we plot $B^{13}$ as a function of $p$ and $w$ in Fig.\ref{1w13}. Comparing with Eq.(\ref{Cc1}) where weak measurement is not applied, we see that the technique of weak measurement successfully preserves the non-bilocal correlation for most of the region of the parameters $p$ and $w$. 

\subsubsection{Case-\ref{d2}}

The joint state among Alice, Bob and Charlie, i.e., $\rho_{AB}^{r} \otimes \rho_{BC}^{r}$ ($\sigma_{AB}^{r} \otimes \sigma_{BC}^{r}$) and the optimal measurement settings for Alice and Charlie used to obtain Eq.(\ref{Cc2}), i.e., in the absence of weak measurement, give rise to the left hand side of Eq.(\ref{1-3}) given by
\begin{align}
B^{13} = & \sqrt{\frac{2(1-p)}{3-2p}} ~ \frac{1}{2-[1+p(1-w)] r -w} \nonumber\\
& \times \Big([1-p(1-w)](2-r)-w \nonumber\\
& + \sqrt{(1-w)(1-r)}\Big)
\end{align}

Maximization of the above expression w.r.t. reverse weak strength $r$ chosen by Alice and Charlie specifies the optimum strength of $r$ ,i.e. $r_{opt} = \frac{1}{(1+p-pw)^2} \Big( w+ p(1-w)[(2+w)+4y(1-w)+8p^2 (2-p)(1-w)^2 +2p(1-w)(3-4w)]\Big)$, where $y=(1-p)\sqrt{(1-p)+p(1-p)(1-w)+4p^2 (1-p)^2 (1-w)^2}$. Hence, using $r=r_{opt}$, we plot $B^{13}$ as a function of $p$ and $w$ in comparison to the expression given by Eq.(\ref{Cc2}). It is observed in Fig.\ref{2w13} that the technique of weak measurement helps to preserve the non-bilocal correlation for a notable range of $p$ and $w$. 
 
\section{Improvement in the average non-bilocal correlations} \label{6}

\begin{figure*}
   \hrulefill
   \begin{subfigure}{6cm}
    \centering\includegraphics[width=5cm]{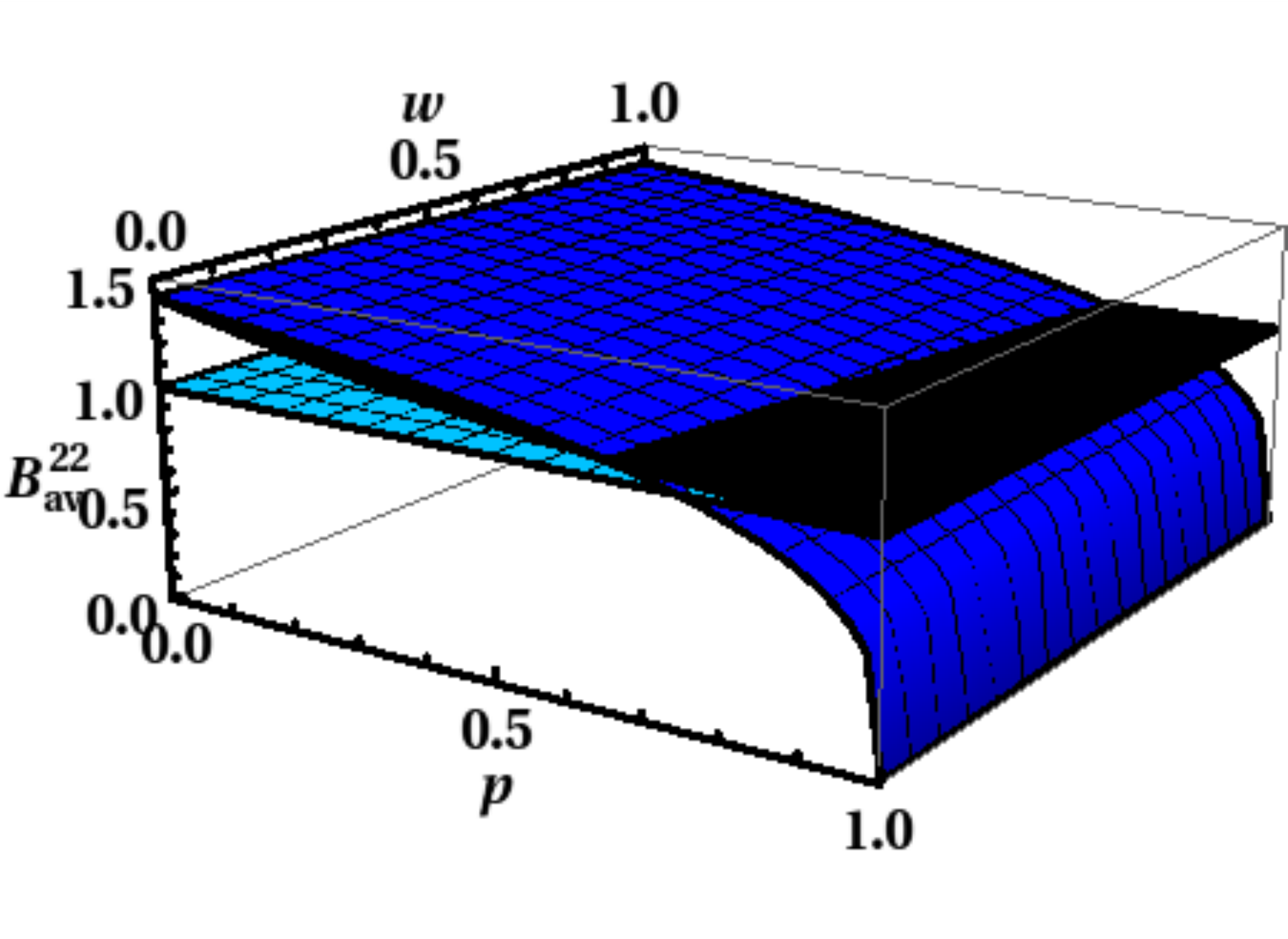}
    \caption{\footnotesize $B^{22}_{av}$ corresponding to case-\ref{d1}}
    \label{avg1}
  \end{subfigure}%
  \begin{subfigure}{6cm}
    \centering\includegraphics[width=5cm]{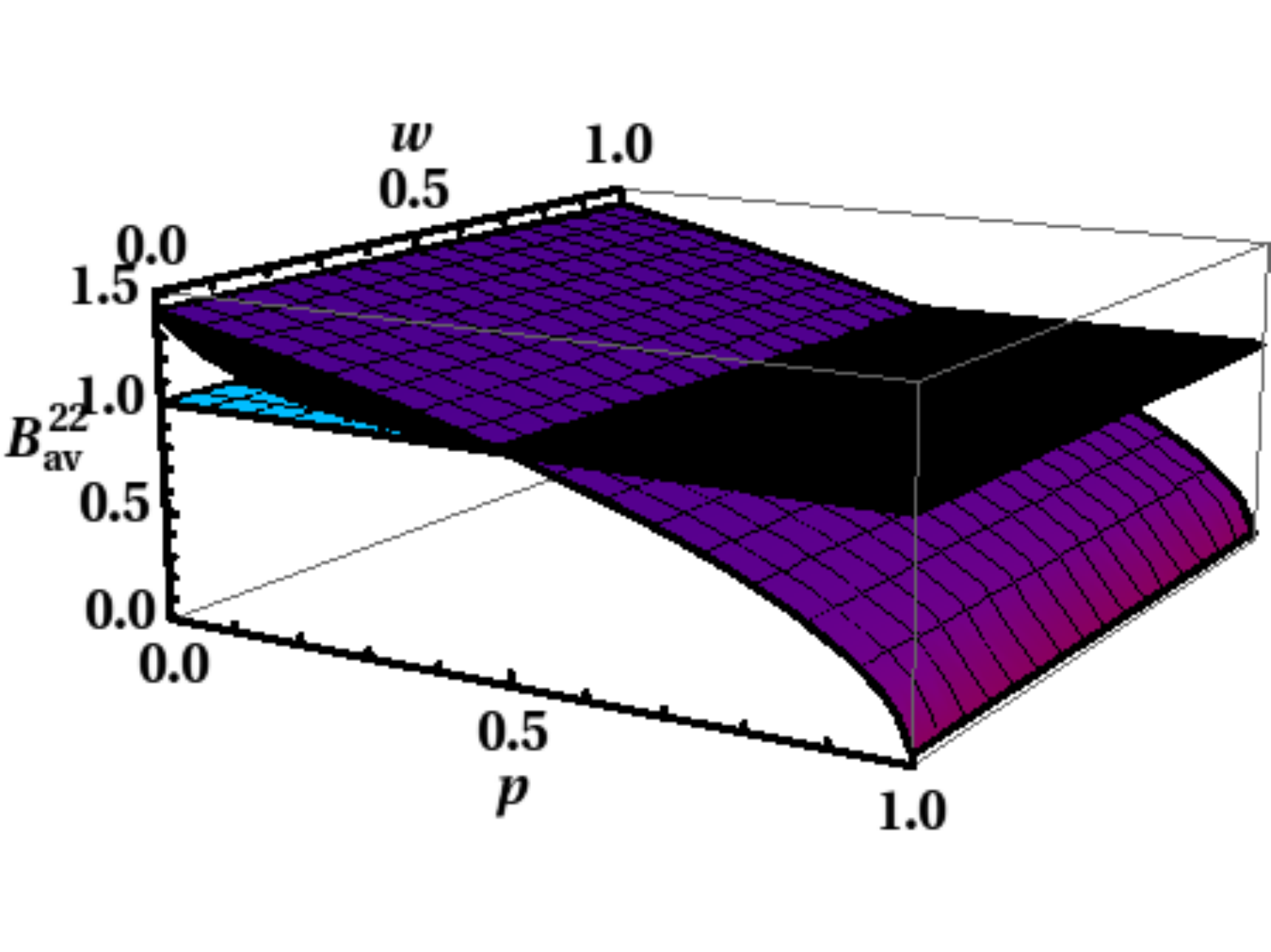}
    \caption{\footnotesize $B^{22}_{av}$ corresponding to case-\ref{d2}}
    \label{avg2}
  \end{subfigure}%
   \begin{subfigure}{6cm}
    \centering\includegraphics[width=5cm]{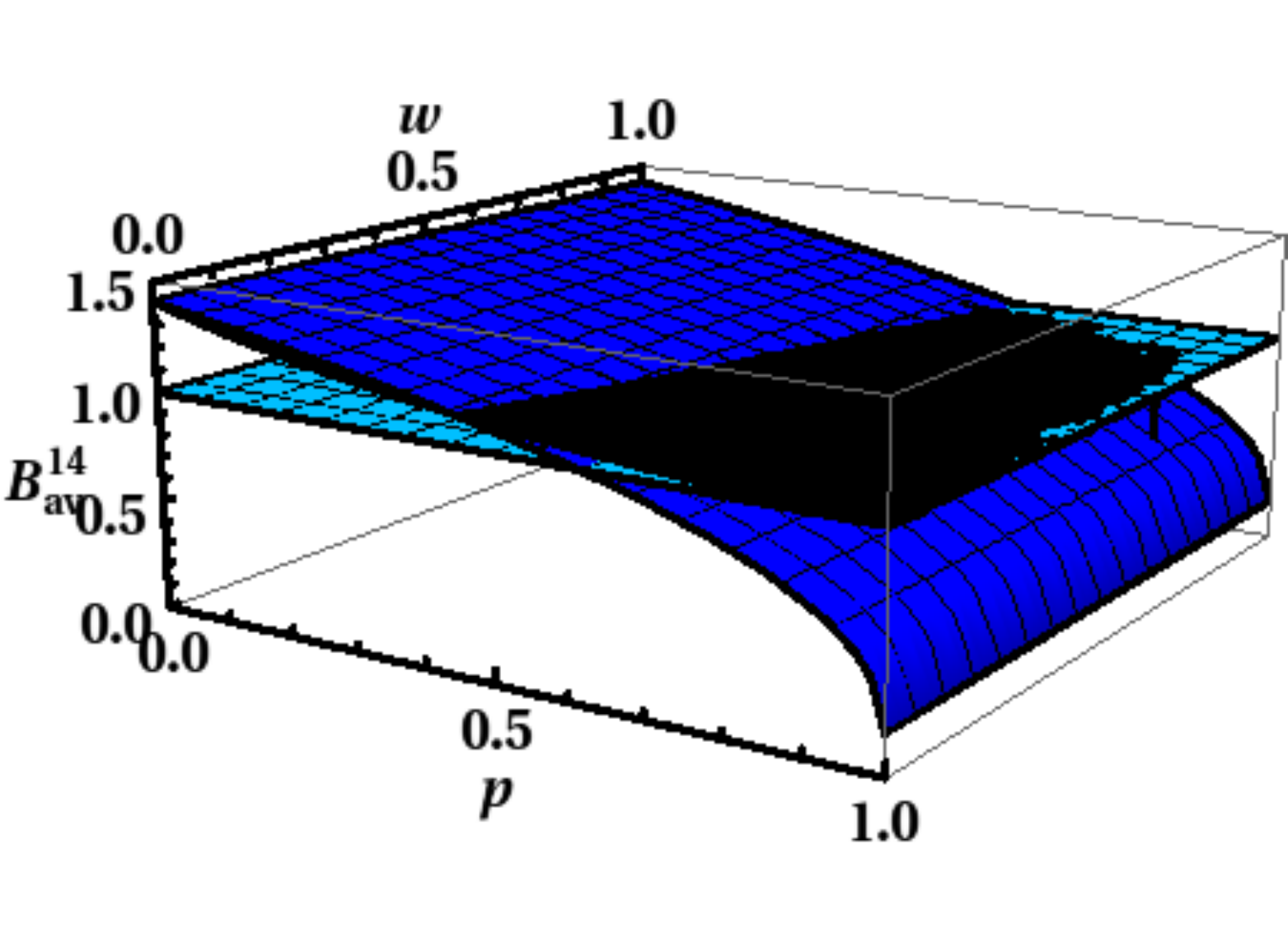}
    \caption{\footnotesize $B^{14}_{av}$ corresponding to case-\ref{d1}}
    \label{avg3}
  \end{subfigure}
  
  \begin{subfigure}{6cm}
    \centering\includegraphics[width=5cm]{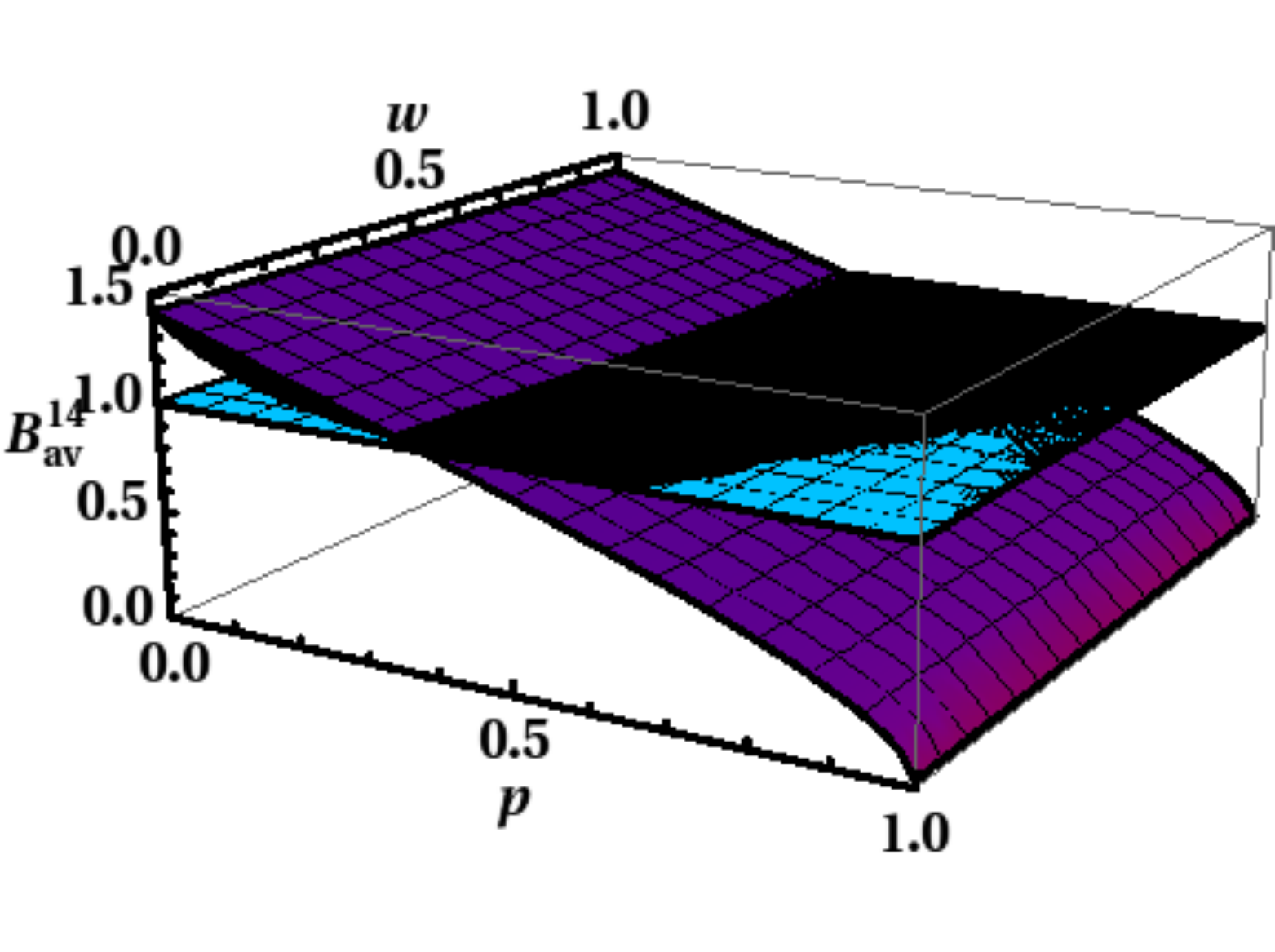}
    \caption{\footnotesize $B^{14}_{av}$ corresponding to case-\ref{d2}}
    \label{avg4}
  \end{subfigure}%
  \begin{subfigure}{6cm}
    \centering\includegraphics[width=5cm]{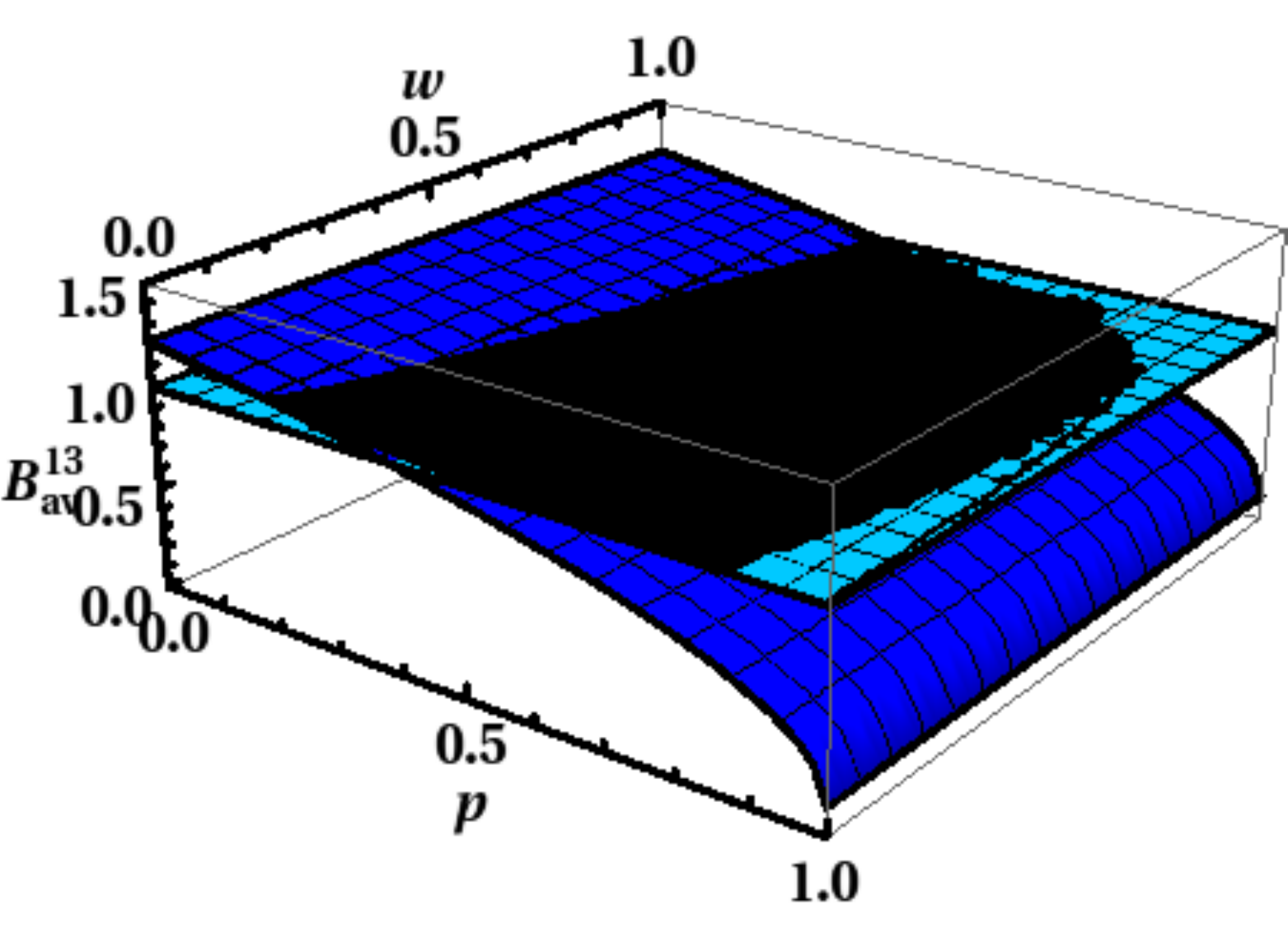}
    \caption{\footnotesize $B^{13}_{av}$ corresponding to case-\ref{d1}}
    \label{avg5}
  \end{subfigure}%
  \begin{subfigure}{6cm}
    \centering\includegraphics[width=5cm]{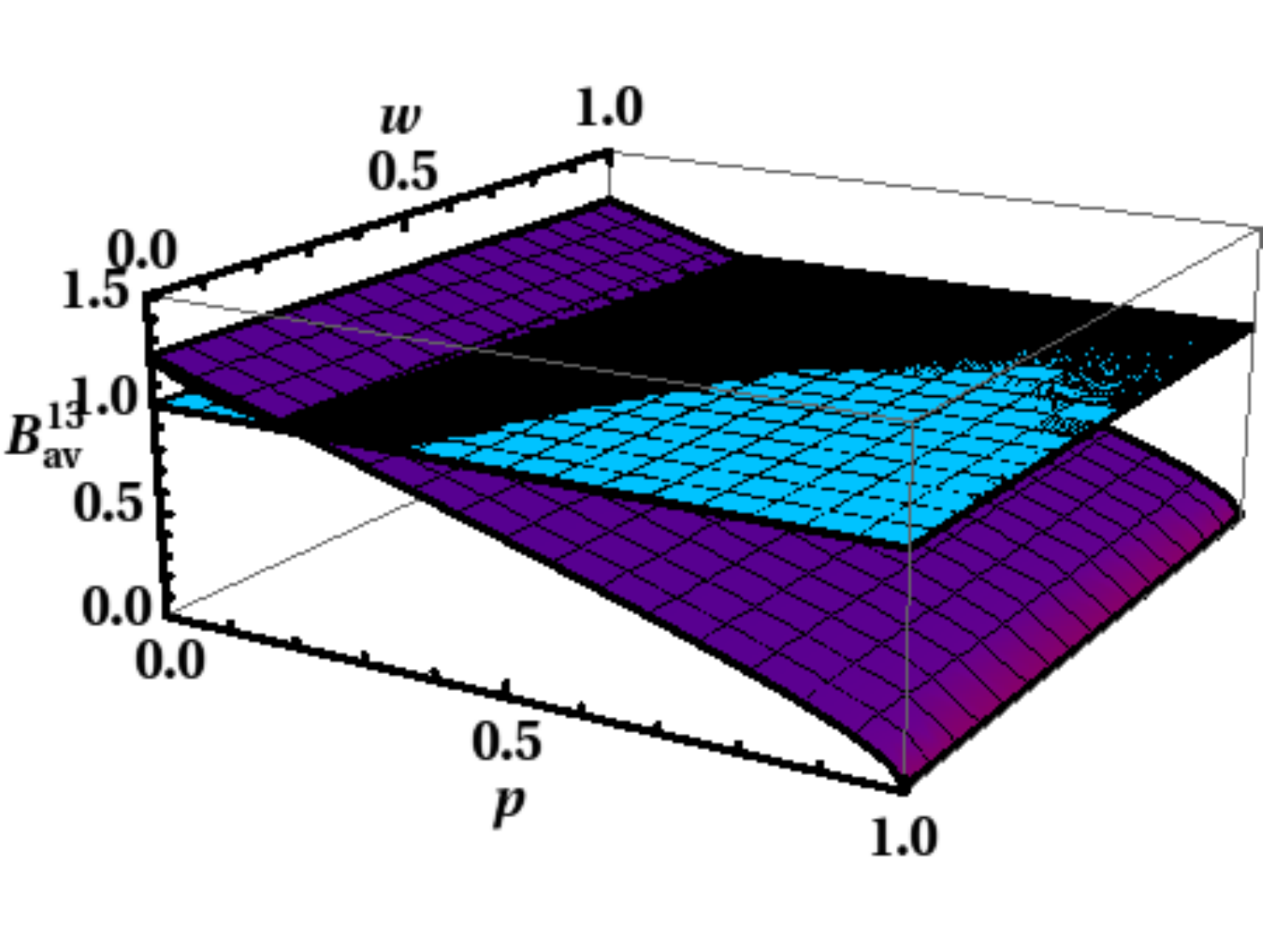}
    \caption{\footnotesize $B^{13}_{av}$ corresponding to case-\ref{d2}}
    \label{avg6}
  \end{subfigure}
  \caption{\footnotesize (Coloronlile) The average of (non-)bilocal correlations at the optimum value of reverse weak measurement given by Eq.(\ref{average}) (upper black surfaces) are plotted w.r.t. the strength of decoherence $p$ (x-axis) and the strength of weak measurement $w$ (y-axis). The lower surfaces correspond to the respective cases without application of weak measurement. The flat surfaces represent the upper bound of bilocal correlations. Importantly, there can be found certain regions in all the three Bell measurement scenarios where average non-bilocal correlations can be preserved using the weak mesurement technique.}
   \hrulefill
   \label{avg}
\end{figure*}

In order to obtain a quantitative understanding of the technique of weak measurements used to preserve non-bilocal correlations, let  us now introduce an average quantity characterizing the average non-bilocal correlation taking into  effect  both the probabilistic success and failure of post-selection in attaining the final state. Corresponding to case-\ref{d1}, the success probability of achieving the final state $\rho_{AB}^{r} \otimes |\psi^{\pm}\rangle_{BC}\langle\psi^{\pm}|$ ($\sigma_{AB}^{r} \otimes |\phi^{\pm}\rangle_{BC}\langle\phi^{\pm}|$) is denoted as $P_{succ}^{r}|_{r=r_{opt}}$, and corresponding to case-\ref{d2}, the success probability of achieving the final state $\rho_{AB}^{r} \otimes \rho_{BC}^{r}$ ($\sigma_{AB}^{r} \otimes \sigma_{BC}^{r}$) is denoted as $P_{succ}^{rr}|_{r=r_{opt}}$. It may be relevant to recall that for each post-selection, the success probability implies the negation in detecting the system.  With the left-over probability, the system is detected, breaking its entanglement, and thus, making the system bilocal (corresponding to the classical bound of $1$) irrespective of whichever scenario of BSM one employs.

In context of  the scenarios of binary input-binary output, one input-four outputs (for Bob), and one input-three outputs (for Bob) we define the following three quantities respectively, which we call average of non-bilocal correlations:
\begin{align}
B^{22}_{av} = P_{succ}^{i}|_{r=r_{opt}} ~ B^{22}|_{r=r_{opt}} + (1- P_{succ}^{i}|_{r=r_{opt}}) \nonumber\\
B^{14}_{av} = P_{succ}^{i}|_{r=r_{opt}} ~ B^{14}|_{r=r_{opt}} + (1- P_{succ}^{i}|_{r=r_{opt}}) \nonumber\\
B^{13}_{av} = P_{succ}^{i}|_{r=r_{opt}} ~ B^{13}|_{r=r_{opt}} + (1- P_{succ}^{i}|_{r=r_{opt}})
\label{average}
\end{align}
where, the superscript $i$ represents the indices $r$ and $rr$ while we compute $\lbrace B^{22},B^{14},B^{13}\rbrace$ corresponding to case-\ref{d1} and case-\ref{d2} respectively.

In Fig.(\ref{avg1}-\ref{avg6})  we plot the quantities $\lbrace B^{22}_{av},B^{14}_{av},B^{13}_{av}\rbrace$ w.r.t. $p$ and $w$ for case-\ref{d1} and case-\ref{d2}, respectively for the three BSM scenarios comparing with the corresponding quantities in the absence of weak measurement. It is observed that though, as expected, the improvement in non-bilocal correlations is not up to the level as obtained through the exclusively successful post-selection (i.e., by discarding detection of the system), the technique of weak measurement is useful in preserving the  average  quantum non-bilocal correlations  for significant domains in the ($p-w$) parametric space for all the three BSM scenarios.

\section{Conclusion} \label{7}

In the present work we investigate the bilocal correlations \cite{cyril,cyril1} involving three parties under the effect of decoherence modelled by the amplitude damping channel. We consider three different scenarios of Bell-state measurement employed by the central party in a procedure of entanglement swapping. Quantum non-bilocal correlations are manifested through the violation of the respective bilocality inequality in a given scenario. We first show that the non-bilocal correlations decrease monotonically under amplitude damping, thus exhibiting their difference with the correlations responsible for performing teleportation \cite{except2,except3}. The above result can be easily anticipated since there is no underlying role of classical correlation in enhancing non-bilocal correlation. We then embark on the main goal of the paper, {\it viz.}, studying the effect of
weak measurement in preserving such non-bilocal correlations. 

We employ the technique of weak measurement and its reversal which has been known to preserve certain other types of quantum correlations \cite{weak1, weak2, weak3} against the effect of decoherence. Here, weak measurement is performed by the central party, which helps to slow down the effect of amplitude damping on the state when the particles are transmitted to the two other parties. Reverse weak measurements are then performed by either or both the other parties with optimum strength required for maximizing the final non-bilocal correlations. We show that irrespective of the choice of the initial Bell state, this procedure enables preservation of the non-bilocal correlations for a large range of parameters. We consider three different types of Bell state measurements to show that this technique of protecting quantum correlations against decoherence through weak measurements works generically, though with certain quantitative differences in the magnitude of correlations preserved. The average strength of non-bilocal correlations is enhanced even if one takes into account the success probability of the post-selection process involved in weak measurements. Before concluding, we note that further such studies involving other decoherence channels might be worthwhile in order to ascertain the practical applicability of such non-bilocal correlations in recently suggested quantum communications protocols \cite{expt2,carvacho}.

{\it Acknowledgements:} SG thanks Cyril Branciard and Kaushiki Mukherjee for helpful discussions. SD acknowledges support from DST INSPIRE Fellowship, Govt. of India (Grant No. C/5576/IFD/2015-16).

\bibliography{BILOC}

\end{document}